\documentclass[conference]{IEEEtran}
\IEEEoverridecommandlockouts

\usepackage{cite}
\usepackage{amsmath,amssymb,amsfonts}
\usepackage{algorithmic}
\usepackage{graphicx}
\usepackage{textcomp}
\usepackage{xcolor}
\usepackage{amsmath} 
\usepackage{booktabs} 
\usepackage{array}
\usepackage{graphicx} 
\usepackage{multirow}
\usepackage{color}
\definecolor{Violet}{RGB}{148,0,211} 
\usepackage{makecell} 
\usepackage{float}
\usepackage{subcaption}
\usepackage{colortbl}
\usepackage{placeins}
\usepackage[linesnumbered,vlined,ruled,noend]{algorithm2e}
\usepackage{comment}
\usepackage{pifont}
\usepackage{url} 
\usepackage[table]{xcolor}
\usepackage{listings}
\usepackage{hyperref} 

\newcommand{\cmark}{\ding{51}} 
\newcommand{\xmark}{\ding{55}} 

\def\BibTeX{{\rm B\kern-.05em{\sc i\kern-.025em b}\kern-.08em
    T\kern-.1667em\lower.7ex\hbox{E}\kern-.125emX}}

\makeatletter
\newcommand{\linebreakand}{%
  \end{@IEEEauthorhalign}
  \hfill\mbox{}\par
  \mbox{}\hfill\begin{@IEEEauthorhalign}
}
\makeatother

\begin{document}

\title{Rethinking Text-to-SQL: Dynamic Multi-turn SQL Interaction for Real-world Database Exploration}

\author{
\IEEEauthorblockN{
Linzhuang Sun$^{1\dagger}$,
Tianyu Guo$^{2\dagger}$,
Hao Liang$^{2\dagger}$,
Yuying Li$^{3}$,
Qifeng Cai$^{2}$,
Jingxuan Wei$^{1}$,
Bihui Yu$^{1}$,\\
Wentao Zhang$^{2*}$,
Bin Cui$^{2*}$
}
\IEEEauthorblockA{
\textit{$^{1}$ University of Chinese Academy of Sciences, Beijing, China} \\
\textit{$^{2}$ Peking University, Beijing, China}
\textit{$^{3}$ Tsinghua University, Beijing, China} 
}
\IEEEauthorblockA{
sunlinzhuang21@mails.ucas.ac.cn, \{tianyu.guo, hao.liang\}@stu.pku.edu.cn\\
\{wentao.zhang, bin.cui\}@pku.edu.cn
}
}

\maketitle
\renewcommand{\thefootnote}{\fnsymbol{footnote}}
\footnotetext[2]{Equal contribution.}
\footnotetext[1]{Corresponding authors.}
\renewcommand{\thefootnote}{\arabic{footnote}}

\begin{abstract}
Recent progress in Text-to-SQL research has led to remarkable performance in static, single-turn settings, where models generate SQL queries from natural language questions through multi-step reasoning. However, such systems remain inadequate for real-world, interactive applications, where user intents evolve dynamically and queries must be refined over multiple turns. In practical scenarios such as financial or business analytics, users iteratively modify query constraints or dimensions based on intermediate results. 
To systematically evaluate these capabilities, we introduce DySQL-Bench, a benchmark designed to assess model's ability under dynamic user interactions. Unlike previous manually curated datasets, DySQL-Bench is constructed through an automated two-stage pipeline consisting of task synthesis and task verification. In the synthesis stage, raw database tables are transformed into structured tree representations, each capturing the logical relationships within a complete user record. These structured trees serve as the foundation from which Large Language Models (LLMs) generate diverse and realistic evaluation tasks. In the verification stage, an interaction-oriented quality control module filters out erroneous samples, followed by expert validation to ensure data reliability. Through rigorous human evaluation, our automatic synthesis data achieves 100\% correctness.
We further propose a multi-turn dynamic evaluation framework simulating a realistic interaction environment involving three roles: an LLM-simulated user, the model under evaluation, and an executable database system. The simulated user interacts with the model based on task instructions and intermediate responses, while the model must adapt its reasoning strategy and SQL generation in response to evolving user intents.
DySQL-Bench integrates databases from BIRD and Spider 2, spanning 13 domains and comprising 1,072 evaluation tasks. Experimental results demonstrate that even advanced models such as GPT-4o achieve only 58.34\% overall accuracy and 23.81\% on the Pass\^{}5 metric, highlighting the significant challenge posed by our benchmark. All code and benchmark data are publicly available at https://github.com/Aurora-slz/Real-World-SQL-Bench.
\end{abstract}


\section{Introduction}\label{sec:intro}
Structured Query Language (SQL) has become a cornerstone of data-driven applications across domains such as customer service analytics, electronic health record exploration, and financial risk monitoring~\cite{katsogiannis2023survey, shi2025survey}. For domain experts with clearly defined analytical objectives, a single, well-crafted SQL query is often sufficient to obtain the desired information. However, in many real-world scenarios, users cannot fully articulate their intentions in a single turn due to incomplete prior knowledge, ambiguous goals, or evolving analytical requirements~\cite{zhang2025survey}. As a result, interactions with databases naturally unfold as multi-turn dialogues, where users iteratively refine query constraints, clarify ambiguities, and progressively explore the data space.

\begin{figure*}[t]
\centering
\includegraphics[width=0.85\textwidth]{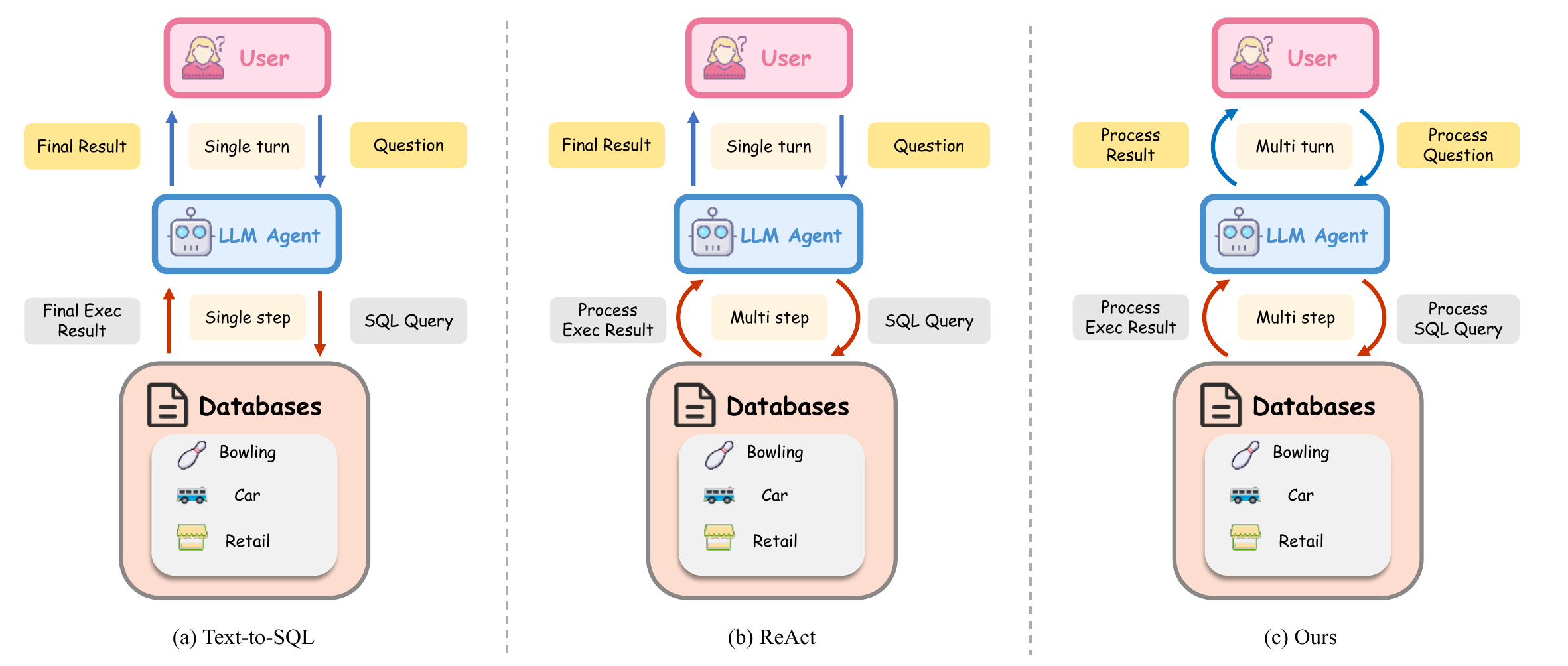}
\caption{Overview of interaction types between user, LLM agent, and database.
From left to right: (a) direct Text-to-SQL execution, (b) iterative reasoning within a single query (ReAct), and (c) our approach enabling multi-step, multi-turn contextual execution.}
\label{fig:demo}
\end{figure*}

\begin{table*}[htbp]
  \centering
  \caption{Comparison of Text2SQL datasets across key dimensions including construction method, CRUD operation support, evaluation paradigm, and task complexity in both single-turn and multi-turn settings.}
  \resizebox{1.0\linewidth}{!}{
    \begin{tabular}{c|c|c|c|cc|cc}
    \toprule
    \multirow{2}[2]{*}{\textbf{Dataset}} & \multirow{2}[2]{*}{\textbf{Construction}} & \multirow{2}[2]{*}{\textbf{CRUD}} & \multirow{2}[2]{*}{\textbf{Evaluation}} & \multicolumn{2}{c}{\textbf{Task Single-Turn}} & \multicolumn{2}{c}{\textbf{Task Multi-Turn}} \\
          &       &       &       & \textbf{Single Step} & \textbf{Multi Step} & \textbf{Single Step} & \textbf{Multi Step} \\
    \midrule
    BIRD  & Human & SELECT & Static & \cmark & \cmark & \xmark & \xmark \\
     CoSQL & Human & SELECT & Static & \cmark & \cmark & \xmark & \xmark \\
    MultiSQL & LLM-Gen + Human & FULL  & Static & \cmark & \cmark & \cmark & \cmark \\
    SPIDER & Human & SELECT & Static & \cmark & \cmark & \xmark & \xmark \\
    SPIDER 2.0 & Human & SELECT & Static & \cmark & \cmark & \xmark & \xmark \\
    \midrule
    \textbf{DySQL-Bench} & \textbf{LLM-Gen + Human} & \textbf{FULL} & \textbf{Dynamic} & \textbf{\cmark} & \textbf{\cmark} & \textbf{\cmark} & \textbf{\cmark} \\
    \bottomrule
    \end{tabular}%
    }
  \label{tab:bench_compare}%
\end{table*}%


Recent advances in LLMs have markedly improved performance on Text-to-SQL~\cite{li2025omnisql, ma2025sql, pourreza2025reasoning}. To assess these capabilities, several benchmarks have been introduced, including Spider 1~\cite{yu2018spider}, Spider 2~\cite{lei2024spider}, and BIRD~\cite{li2023can}. Despite their impact, important gaps remain:

\textbf{(1) Limited multi-turn coverage.}
As shown in Figure~\ref{fig:demo}, most datasets and benchmarks focus on single-turn tasks or static multi-turn tasks~\cite{yu2019cosql} and fail to capture realistic conversational workflows in which users iteratively refine queries, correct mistakes, or add constraints across turns.

\textbf{(2) Incomplete CRUD spectrum.}
Although MultiSQL~\cite{li2024multisql} is an early effort toward covering the full CRUD spectrum, the majority of existing benchmarks still emphasize read-only $SELECT$ queries and provide little to no assessment of write operations, $INSERT/UPDATE/DELETE$, which frequently arise in practice.

\textbf{(3) Lack of dynamic multi-turn evaluation protocols.}
Even when multi-turn data are available, there is no standardized evaluation protocol to handle stateful execution, measure behavior consistency, or account for evolving database states across turns.


To address this work, we aim to bridge this gap by introducing \textbf{DySQL-Bench}, the first benchmark explicitly designed to evaluate models in \textit{dynamic, multi-turn Text-to-SQL} scenarios. As shown in Table~\ref{tab:bench_compare}, our benchmark measures a model’s ability to perform contextually grounded SQL reasoning that integrates the full spectrum of \textbf{CRUD (Create, Read, Update, Delete)} operations, thereby reflecting the genuine demands of real-world data exploration and management. To construct this benchmark efficiently and reproducibly, we design a \textit{fully automated two-stage data synthesis pipeline}. In the first stage, raw database tables are transformed into \textit{tree-structured representations} that capture inter-table logical relationships, serving as a structured source pool for task generation. In the second stage, an \textit{interaction-oriented quality control} module filters and validates LLM-generated tasks, followed by human expert verification to ensure semantic accuracy and execution consistency.

Furthermore, we propose a \textit{multi-turn evaluation framework} that simulates realistic user--model--database interaction. In this setting, an LLM-simulated user issues evolving instructions, the model under evaluation generates and executes SQL queries, and an executable database provides intermediate feedback. This interactive setup enables the assessment of how well a model can maintain conversation state, recover from execution errors, and refine its reasoning strategy over time.

In summary, our contributions are as follows:
\begin{itemize}
    \item \textbf{DySQL-Bench}, a new large-scale benchmark for dynamic, multi-turn Text-to-SQL tasks covering the full CRUD spectrum.
    \item \textbf{Two-Stage Automatic Task Synthesis Pipeline}: With the two stage pipeline, we construct high-quality interaction tasks from real databases with minimal manual effort. The pipeline is proven to be effective, through rigorous human evaluation, our automatic synthesis data achieves 100\% correctness.
    \item \textbf{User-Model-Database Evaluation Framework} This enables systematic assessment of contextual reasoning, adaptability, and error recovery across multi-turn interactions.
\end{itemize}


\begin{figure*}[t]
\centering
\includegraphics[width=1.0\textwidth]{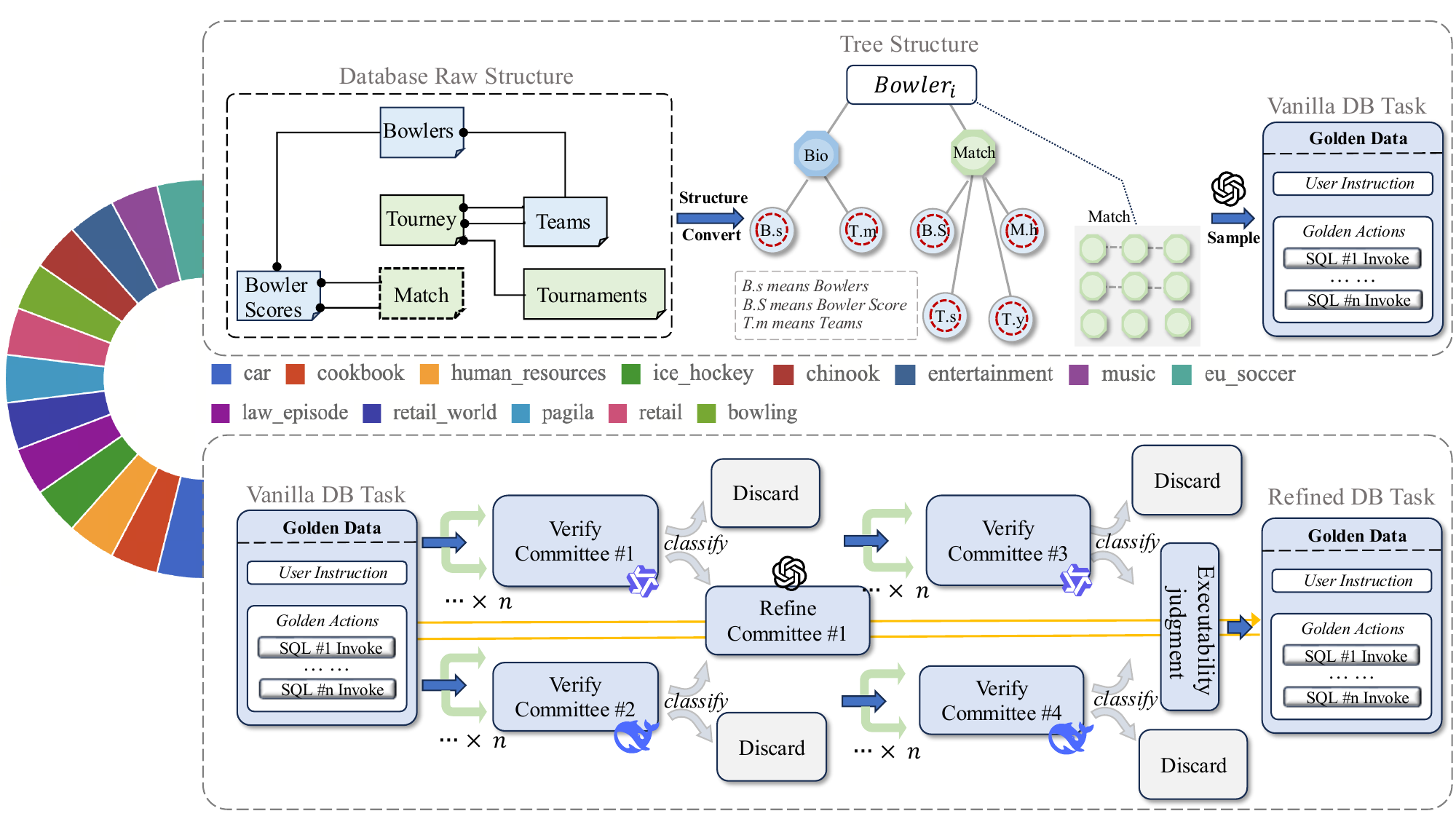}
\caption{Task Generation Pipeline. Based on 13 databases from Spider 2 and BIRD, we designed a two-stage DynSQL-Bench task synthesis method. In the first stage, an LLM generates initial tasks. In the second stage, a filtering mechanism selects high-quality tasks.}
\label{fig:method}
\end{figure*}

\section{Related Work}

\subsection{Text-to-SQL.} Mapping natural language utterances into executable SQL queries has long been viewed as a promising way to democratize database access, freeing users from the burden of mastering schema intricacies and SQL syntax~\cite{qin2022survey, katsogiannis2023survey}. Recent progress in LLMs~\cite{an2025unictokens, an2024mc, luo2024llm, lin2025perceiveanythingrecognizeexplain, lin2024draw} has substantially advanced this direction, driven by their powerful reasoning and cross-domain generalization abilities~\cite{guo2025deepseek, yang2025qwen3, chen2025learning}.
A number of recent efforts have sought to refine this paradigm by decomposing the problem and leveraging contextual reasoning~\cite{sun2024beats, li2025swe}. For instance, few-shot frameworks such as DIN-SQL~\cite{pourreza2023din} and DAIL-SQL~\cite{gao2023text} employ in-context demonstrations to separate schema linking from SQL generation, while DTS-SQL~\cite{pourreza2024dts} enhance smaller-scale models through selective data curation.
In parallel, agent-style systems that integrate thought, action, and feedback, like MAC-SQL~\cite{wang2023mac}, illustrating that iterative interaction with the environment can lead to notable performance gains~\cite{yao2022react}.
Although these studies have collectively improved SQL synthesis accuracy, the majority of them remain confined to single-turn settings~\cite{hong2025next}. Consequently, their robustness and adaptability in multi-turn conversational scenarios have yet to be systematically explored.

\subsection{Text-to-SQL Benchmarks.} Benchmark development has been central to progress in Text-to-SQL research~\cite{zhang2024benchmarking, mitsopoulou2025analysis, bhaskar2023benchmarking}. 
Early datasets such as ATIS~\cite{price1990evaluation}, a flight-booking system dataset mapping natural‐language user queries about airline travel into structured queries, and GeoQuery~\cite{zelle1996learning}, a U.S. geography question–answering dataset converting natural-language questions into formal queries, provided domain-specific testbeds that enabled early system design but lacked schema diversity and compositional depth.
The introduction of SPIDER~\cite{yu2018spider} fundamentally transformed the field by emphasizing cross-domain generalization to unseen databases, catalyzing advances in schema linking, compositional reasoning, and data augmentation. Building on this foundation, follow-up benchmarks like SPIDER 2.0~\cite{lei2024spider} and BIRD~\cite{li2023can} introduced richer database schemas, paraphrased queries, and dynamic evaluation settings to approximate realistic database interaction.
However, existing benchmarks remain largely confined to single-turn query formulation, assuming that a user’s intent can be fully captured in one utterance. In real-world scenarios, users often express goals progressively, refining or expanding their requests based on intermediate outcomes. 

\subsection{Multi-turn Text-to-SQL.} In real-world applications, user queries are often ambiguous, incomplete, or evolve through conversation~\cite{yao2024tau, barres2025tau}. Multi-turn Text-to-SQL research thus focuses on handling underspecified queries by leveraging clarification and context tracking. Early datasets such as CoSQL~\cite{yu2019cosql} extend the SPIDER~\cite{yu2018spider} benchmark with dialogue-based turns to simulate this process.
However, these benchmarks assume static and noise-free dialogue histories, neglecting that systems may initiate different clarification strategies~\cite{yi2024survey}. More recent works investigating autonomous agents that maintain dynamic conversational states~\cite{yao2024tau, li2025beyond}, yet their methodologies have not been adapted to the Text-to-SQL context.
Constructing an effective user simulator for this task remains non-trivial: it must balance database realism with controlled answer spaces and schema constraints.
To bridge this gap, our study introduces a multi-turn benchmark featuring an user simulator, dynamic evaluation and real-world databases. This framework enables a systematic evaluation of reasoning-oriented models under realistic and uncertain Text-to-SQL conditions.

\begin{figure}
    \centering
    \includegraphics[width=\linewidth]{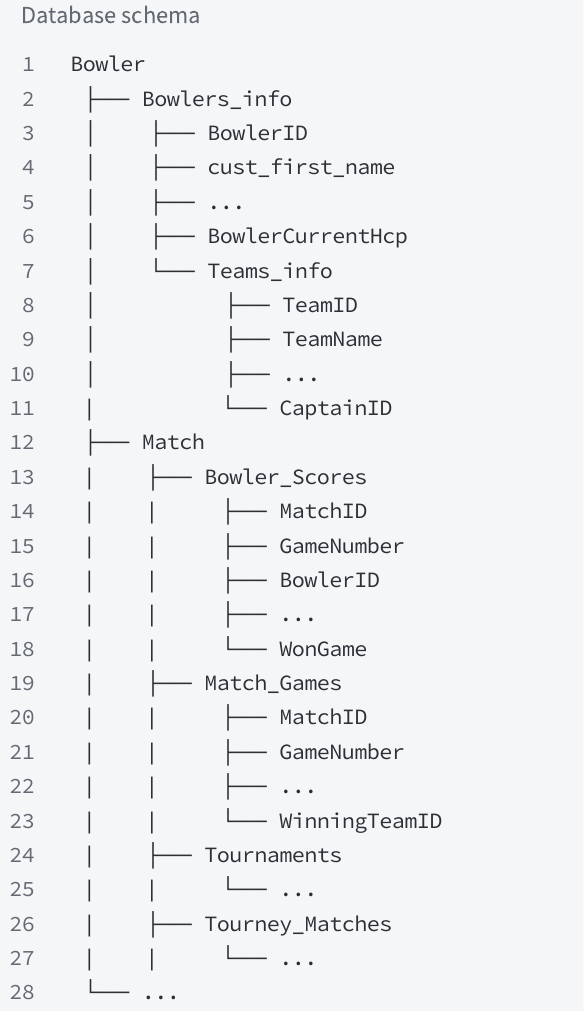}
    \caption{Database Schema Representation.}
    \label{fig:info tree}
\end{figure}

\section{Task Definition}
In this paper, we introduce a multi-turn Text-to-SQL task designed to capture dynamic user behaviors. In this task, a LLM is assigned the role of a simulated user, initialized with an instruction $I$ that specifies both the user’s information and its interaction style. The simulated user then engages in a dialogue with a SQL model. During the interaction, the SQL model generates a sequence of SQL operations $A = (a_1, a_2, \ldots, a_n)$ to manipulate the underlying database $S^1$, resulting in an updated database state $S^2$.

To evaluate whether the generated action sequence $A$ successfully satisfies the user’s request, we construct for each task a golden action list $A_g = (a_1^g, a_2^g, \ldots, a_m^g)$ through a combination of automated synthesis and manual validation. Executing $A_g$ on the initial database state $S^1$ yields a golden database state $S^3 = \text{execute}(A_g \mid S^1)$. 
To objectively determine whether the model-generated action sequence $A$ leads to the same final database state as $A_g$, we compute a hash value for the database states $S^2$ and $S^3$. Specifically, all tables in the database are stored in a predefined order, and the data from each table are retrieved sequentially and serialized into a unified structure. Columns related to update or creation timestamps (e.g., \textit{updated\_at}, \textit{created\_at}, \textit{timestamp}) are excluded to eliminate volatility caused by execution time differences. The serialized data are then converted into a hash value using a consistent hashing function. After executing both the model-generated action sequence $A$ and the golden sequence $A_g$, on identical initial database copies, we compare their resulting hash values. If the two hash values are identical, we consider the task successfully solved—indicating that the model’s execution produced a database state equivalent to the golden reference. Otherwise, the task is marked as unsolved.


\section{Multi-turn DB Tasks Generation}

To ensure the authenticity and complexity of our tasks, our benchmark is based on 13 datasets provided by BIRD and SPIDER2. The construction of benchmark tasks consists of two stages: vanilla task synthesis and task correctness verification.

\subsection{Vanilla DB Task Generation}

The construction of DySQL-Bench fundamentally relies on LLM-based data synthesis, which in turn depends on database-driven sampling as its core data acquisition mechanism. However, directly performing sampling through online SQL queries can be inefficient due to the complex and large-scale nature of real-world databases, where frequent SELECT operations incur significant latency. To address this challenge, we transform the original relational database schema into a designed hierarchical tree structure, serialized in JSON format. This structure allows the system to sample data efficiently and generate tasks for LLMs without repeated database queries.
The construction of the hierarchical structure follows two key steps:

\begin{itemize}

\item \textbf{Primary Table Identification and Root Node Construction:} We first conduct a domain-specific analysis of the database to identify a primary table. Each record in this table represents a core information entity and serves as the root node of the tree.

\item \textbf{Foreign Key Traversal and Hierarchical Expansion:} Based on the foreign key relationships in the primary table, we recursively retrieve all associated records from related tables. These linked entities are added as child nodes, forming a complete hierarchical tree that captures the full relational context of each primary entity.

\end{itemize}

In practical, we selected 13 domains databases from BIRD and SPIDER2. 
For example, in $bowling$ database, we set the $Bowlers$ table as the primary table (Figure~\ref{fig:method}). For each $Bowler_i$ in $Bowlers$ table, we then identify all related information from associated tables and organize these connections into a tree structure, where the primary table serves as the root node and linked tables populate the child nodes, as shown in 
Figure~\ref{fig:info tree}.

Next, we will assign GPT-4.1 as the task generator. Guided by the prompt and using the tree-structured information as the source data, it will generate user instruction tasks $I$ along with their corresponding standard actions $A$.

\subsection{Refined DB Task}

\begin{table*}[htbp]
  \centering
  \caption{Bench Statistics. Abbreviations: BO = bowling, CA = car, CH = chinook, CK = cookbook, EN = entertainment, ES = eu\_soccer, HR = human\_resources, IH = ice\_hockey, LE = law\_episode, MU = music, PA = pagila, RE = retail, RW = retail\_world.}
  \resizebox{1.0\linewidth}{!}{
    \begin{tabular}{lccc|cccccc|cccc|c}
    \toprule
          & \multicolumn{3}{c|}{\textbf{Sports Domain}} & \multicolumn{6}{c|}{\textbf{Entertainment Domain}} & \multicolumn{4}{c|}{\textbf{Business Domain}} & \multirow{2}[4]{*}{\textbf{ALL}} \\
\cmidrule{2-14}          & \textbf{ES} & \textbf{IH} & \textbf{BO} & \textbf{EN} & \textbf{MU} & \textbf{LE} & \textbf{CK} & \textbf{CH} & \textbf{PA} & \textbf{CA} & \textbf{HR} & \textbf{RE} & \textbf{RW} &  \\
    \midrule
    \rowcolor[rgb]{ .867,  .922,  .969} \multicolumn{15}{c}{\textit{\textbf{Task Number of Different Difficulty}}} \\
    \midrule
    \textbf{\# Short} & 153 & 16 & 31 & 46 & 4 & 40 & 30 & 20 & 46 & 23 & 24 & 115 & 13 & \textbf{561} \\
    \textbf{\# Long}  & 62  & 12 & 80 & 85 & 17 & 13 & 21 & 24 & 59 & 4  & 16 & 90  & 18 & \textbf{501} \\
    \textbf{\# All}   & 215 & 28 & 111 & 131 & 21 & 53 & 51 & 44 & 105 & 27 & 40 & 205 & 31 & \textbf{1072} \\
    \midrule
    \rowcolor[rgb]{ .867,  .922,  .969} \multicolumn{15}{c}{\textit{\textbf{CRUD Type Ratio in Each Domain (\%)}}} \\
    \midrule
    \textbf{\% SELECT} & 18.01 & 21.52 & 47.96 & 32.86 & 28.95 & 25.69 & 13.33 & 50.31 & 33.77 & 12.28 & 33.33 & 21.08 & 26.67 & \textbf{28.93} \\
    \textbf{\% UPDATE} & 78.68 & 55.70 & 12.96 & 49.05 & 21.05 & 55.96 & 51.11 & 25.15 & 39.61 & 64.91 & 50.00 & 53.73 & 62.22 & \textbf{49.64} \\
    \textbf{\% INSERT} & 1.96  & 22.78 & 23.15 & 12.38 & 21.05 & 13.76 & 21.48 & 6.75  & 1.62  & 15.79 & 12.50 & 8.96  & 8.89  & \textbf{10.63} \\
    \textbf{\% DELETE} & 1.35  & 0.00  & 15.93 & 5.71  & 28.95 & 4.59  & 14.07 & 17.79 & 25.00 & 7.02  & 4.17  & 16.23 & 2.22  & \textbf{10.80} \\
    \bottomrule
    \end{tabular}%
  }
  \label{tab:bench_statistics}%
\end{table*}%

Directly generated tasks $<I, A>$ often suffer from critical issues, including semantic mismatches between user instructions and golden actions, hallucinated attributes that do not exist in the database, and SQL statements containing syntactic errors that render them unexecutable. Therefore, to reduce the cost of manual calibration and improve the accuracy of automated validation, we design a multi-stage data cleaning pipeline.

Particularly, for each vanilla task, validation is first conducted by a Verifier Committee composed of multiple LLM-based validators. Each validator performs $n$ independent checks, and a task is considered verified only if all $n$ checks of two distinct validators, DeepSeek-r1~\cite{guo2025deepseek} and Qwen3-235B-A22B-2507~\cite{yang2025qwen3}. After this stage, we observed that even when golden actions could technically resolve the instruction, the instruction itself sometimes omitted necessary parameters due to limitations of the initial prompting. To address this issue, we introduce a data refinement stage, where parameters required by the golden actions are backfilled into the instruction.
Following refinement, the updated tasks undergo a second round of committee validation to ensure that no additional hallucinations are introduced. After that, We then test the executability of golden actions in a mock runtime environment, discarding all tasks containing SQL syntax errors or execution failures. 

However, while this pipeline effectively eliminates the majority of noisy data, ensuring benchmark rigor requires human oversight. Therefore, we establish a Quality Assurance Board consisting of ten domain experts, who manually inspect each remaining task to confirm that the golden actions faithfully satisfy the user instruction. Only tasks passing this final inspection are included in the benchmark.

\begin{figure}[htbp]
    \centering
    \includegraphics[width=0.45\textwidth]{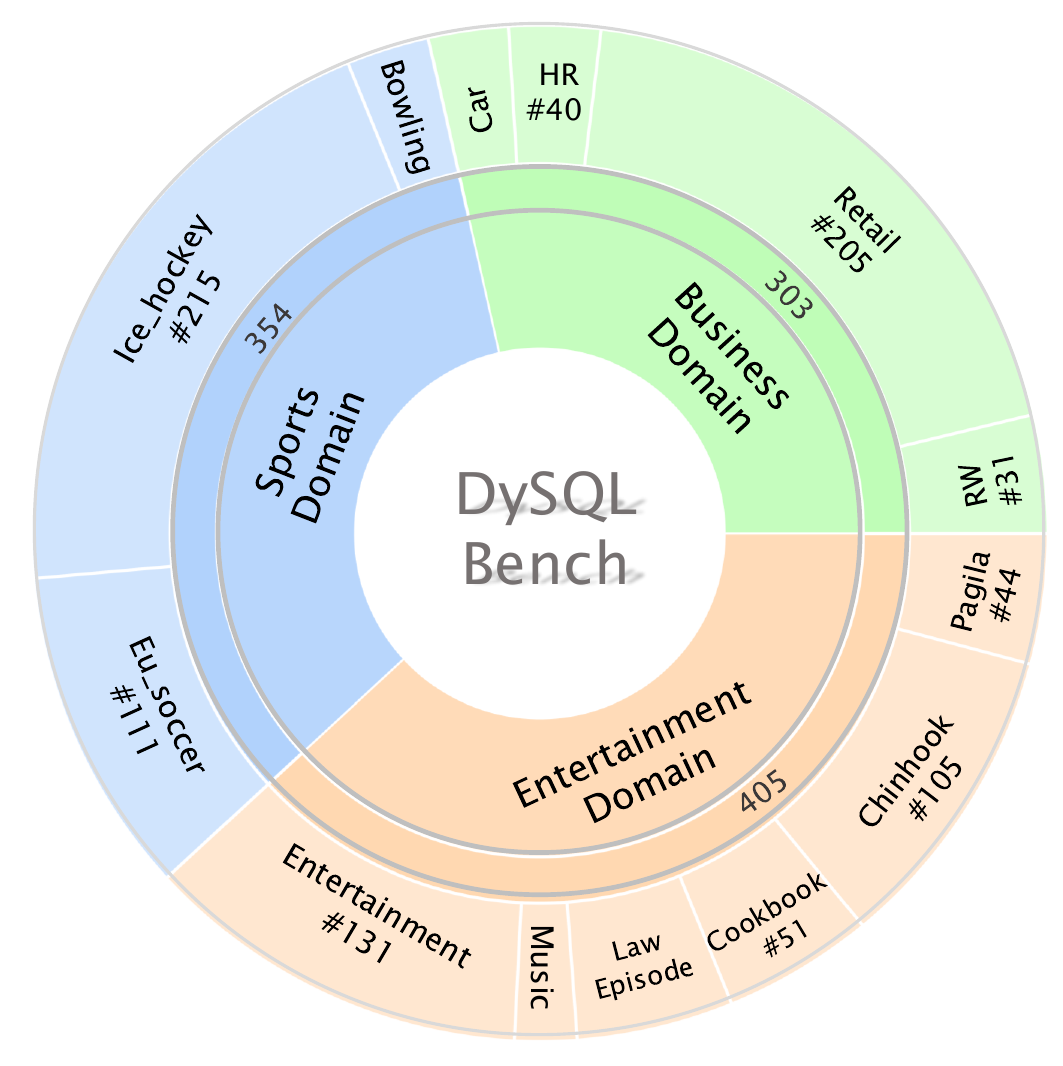}
    \caption{Overview of the DySQL-Bench.}
    \label{fig:bench_bing}
\end{figure}

\subsection{Human Expert Evaluation}
To ensure the reliability and correctness of the benchmark, every task in the final dataset underwent rigorous human verification. Specifically, after passing the multi-stage automatic validation pipeline, all $1{,}072$ tasks were manually inspected by a Quality Assurance Board composed of ten domain experts with extensive experience in database management and SQL semantics.

Each expert independently reviewed the paired instruction–action tuples $\langle I, A \rangle$ to confirm three aspects:
(1) the \textit{semantic fidelity} between the user instruction and the corresponding SQL action;
(2) the \textit{structural validity} of the SQL syntax and its logical coherence with the database schema; and
(3) the \textit{executability and outcome correctness} within the simulated database environment.

Any disagreements among annotators were resolved through cross-review and consensus discussions, ensuring inter-annotator consistency. The committee reported a near-perfect agreement rate ($>99.5\%$ Cohen’s $\kappa$), and all discrepancies were manually corrected before release.

As a result, every one of the 1,072 benchmark tasks has been verified to be semantically correct, executable, and free from hallucinated or ill-formed SQL queries. This full-scope human validation guarantees that the benchmark can serve as a high-fidelity, error-free foundation for evaluating interactive database reasoning and manipulation in large language models.

\begin{table*}[t]
\caption{Short. Performance comparison of open-source and proprietary models on the Real-World-SQL-Bench benchmark. Abbreviations: BO = bowling, CA = car, CH = chinook, CK = cookbook, EN = entertainment, ES = eu\_soccer, HR = human\_resources, IH = ice\_hockey, LE = law\_episode, MU = music, PA = pagila, RE = retail, RW = retail\_world.}
\resizebox{1.0\linewidth}{!}{
\label{tab:main-results-Easy}
\centering
\begin{tabular*}{\textwidth}{@{\extracolsep{\fill}} l| ccc|cccccc|cccc}
\toprule
\multirow{2}[4]{*}{\textbf{Model}} & \multicolumn{3}{c|}{\textbf{Sports Domain}} & \multicolumn{6}{c|}{\textbf{Entertainment Domain}} & \multicolumn{4}{c}{\textbf{Business Domain}} \\
\cmidrule{2-14}          & \textbf{ES} & \textbf{IH} & \textbf{BO} & \textbf{EN} & \textbf{MU} & \textbf{LE} & \textbf{CK} & \textbf{CH} & \textbf{PA} & \textbf{CA} & \textbf{HR} & \textbf{RE} & \textbf{RW} \\
\midrule
\rowcolor[rgb]{.867, .922, .969}
\multicolumn{14}{c}{\textit{\textbf{Pass\^{}1}}} \\
\midrule
GPT-4o & 64.05 & \underline{56.25} & 67.74 & \underline{80.43} & \underline{75.00} & 27.50 & 30.00 & \underline{70.00} & 58.70 & \underline{52.17} & 58.33 & 55.65 & \textbf{84.62} \\
DeepSeek-V3 & 54.90 & 37.50 & 25.81 & 71.74 & 50.00 & 35.00 & 36.67 & 55.00 & 54.35 & 43.48 & 45.83 & 40.87 & \underline{76.92} \\
Gemini2.5-flash & 53.59 & 18.75 & 48.39 & 54.35 & 25.00 & 15.00 & 30.00 & 35.00 & 17.39 & 21.74 & 20.83 & 1.74 & 53.85 \\
Qwen2.5-Max & \textbf{77.78} & \underline{56.25} & \underline{77.42} & \textbf{86.96} & \textbf{100.00} & 42.50 & \textbf{56.67} & 60.00 & \underline{69.57} & \textbf{56.52} & \textbf{83.33} & \textbf{71.30} & \textbf{84.62} \\
Qwen2.5-72B-Instruct & \underline{72.55} & \textbf{68.75} & \textbf{87.10} & \textbf{86.96} & \underline{75.00} & 60.00 & \underline{50.00} & \textbf{85.00} & \textbf{73.91} & \underline{52.17} & 70.83 & \underline{69.57} & \textbf{84.62} \\
Llama3.1-70B-Instruct & 56.86 & \underline{56.25} & 51.61 & \underline{80.43} & 50.00 & \underline{65.00} & 30.00 & \underline{70.00} & 58.70 & 47.83 & \underline{75.00} & 51.30 & \underline{76.92} \\
OmniSQL-32B & 62.75 & 31.25 & 61.29 & 54.35 & 25.00 & \textbf{72.50} & 33.33 & 60.00 & 36.96 & \underline{52.17} & 70.83 & 41.74 & 38.46 \\
Qwen3-32B & 57.52 & 50.00 & 22.58 & 58.70 & \textbf{100.00} & 27.50 & 30.00 & 55.00 & 54.35 & \underline{52.17} & 50.00 & 53.91 & 46.15 \\
\midrule
\rowcolor[rgb]{.867, .922, .969}
\multicolumn{14}{c}{\textit{\textbf{Pass\^{}3}}} \\
\midrule
GPT-4o & 33.99 & \underline{37.50} & 38.71 & 63.04 & \underline{75.00} & 5.00 & 13.33 & 35.00 & 39.13 & 21.74 & 41.67 & 29.57 & \textbf{76.92} \\
DeepSeek-V3 & 36.60 & 18.75 & 16.13 & 36.96 & 25.00 & 17.50 & 10.00 & 15.00 & 15.22 & 21.74 & 20.83 & 22.61 & 53.85 \\
Gemini2.5-flash & 21.57 & 0.00 & 22.58 & 8.70 & 0.00 & 7.50 & 6.67 & 0.00 & 8.70 & 0.00 & 0.00 & 0.00 & 53.85 \\
Qwen2.5-Max & \textbf{56.86} & \textbf{43.75} & \underline{54.84} & 60.87 & \textbf{100.00} & 25.00 & \underline{23.33} & \underline{55.00} & \underline{41.30} & \underline{30.43} & \underline{50.00} & \underline{40.87} & \underline{69.23} \\
Qwen2.5-72B-Instruct & \textbf{56.86} & \textbf{43.75} & \textbf{64.52} & \textbf{71.74} & \underline{75.00} & \textbf{45.00} & \textbf{30.00} & \textbf{65.00} & \textbf{54.35} & \textbf{43.48} & \textbf{58.33} & \textbf{52.17} & \underline{69.23} \\
Llama3.1-70B-Instruct & 30.72 & \underline{37.50} & 32.26 & \underline{67.39} & 25.00 & 32.50 & 6.67 & \underline{55.00} & 34.78 & 21.74 & 45.83 & 24.35 & \textbf{76.92} \\
OmniSQL-32B & \underline{42.48} & 31.25 & 35.48 & 34.78 & 0.00 & \underline{37.50} & 20.00 & 35.00 & 19.57 & 17.39 & 45.83 & 21.74 & 23.08 \\
Qwen3-32B & 30.07 & 25.00 & 3.23 & 28.26 & 50.00 & 7.50 & 3.33 & 35.00 & 19.57 & 26.09 & 20.83 & 30.43 & 46.15 \\
\midrule
\rowcolor[rgb]{.867, .922, .969}
\multicolumn{14}{c}{\textit{\textbf{Pass\^{}5}}} \\
\midrule
GPT-4o & 23.53 & \underline{31.25} & 25.81 & 41.30 & 25.00 & 2.50 & 6.67 & 25.00 & \underline{23.91} & 8.70 & 33.33 & 17.39 & \underline{61.54} \\
DeepSeek-V3 & 22.22 & 12.50 & 12.90 & 23.91 & 0.00 & 10.00 & 10.00 & 10.00 & 10.87 & 8.70 & 16.67 & 15.65 & 38.46 \\
Gemini2.5-flash & 8.50 & 0.00 & 9.68 & 0.00 & 0.00 & 5.00 & 3.33 & 0.00 & 6.52 & 0.00 & 0.00 & 0.00 & 46.15 \\
Qwen2.5-Max & \underline{43.79} & \textbf{37.50} & \underline{38.71} & \underline{50.00} & \textbf{75.00} & 7.50 & \underline{13.33} & \underline{45.00} & \textbf{32.61} & \underline{17.39} & \underline{41.67} & \underline{26.09} & 53.85 \\
Qwen2.5-72B-Instruct & \textbf{49.67} & 31.25 & \textbf{54.84} & \textbf{63.04} & \underline{50.00} & \textbf{32.50} & \textbf{26.67} & \textbf{65.00} & \textbf{32.61} & \textbf{43.48} & \textbf{58.33} & \textbf{35.65} & \textbf{69.23} \\
Llama3.1-70B-Instruct & 16.34 & \underline{31.25} & 29.03 & 47.83 & 25.00 & \underline{22.50} & 6.67 & \underline{45.00} & 21.74 & 8.70 & 37.50 & 15.65 & \textbf{69.23} \\
OmniSQL-32B & 30.72 & \underline{31.25} & 25.81 & 23.91 & 0.00 & \underline{22.50} & \underline{13.33} & 25.00 & 13.04 & 13.04 & 25.00 & 9.57 & 23.08 \\
Qwen3-32B & 17.65 & 0.00 & 3.23 & 19.57 & \underline{50.00} & 5.00 & 3.33 & 15.00 & 10.87 & 8.70 & 8.33 & 17.39 & 46.15 \\
\bottomrule
\end{tabular*}
}
\end{table*}

\subsection{Benchmark Task Statistics}

Our benchmark spans 13 distinct domains, encompassing a total of 1,072 tasks with varying levels of complexity. We categorize each task based on the length of its golden action sequence: tasks with fewer than three actions are labeled as \textit{Short}, while those with three or more actions are labeled as \textit{Long}. Under this criterion, the benchmark contains 561 \textit{Short} tasks and 501 \textit{Long} tasks. 
These domains, ranging from sports domain (Bowling, Ice Hockey and Eu Soccer), business domain (Car, Human Resources and Retail), and entertainment domains (Entertainment, Music, Cookbook, Chinhook, Pagila), collectively reflect the diversity of real-world database applications.

Unlike previous Text-to-SQL benchmarks that focus predominantly on static SELECT-style queries, our dataset systematically covers the full CRUD spectrum—Create (INSERT), Read (SELECT), Update (UPDATE), and Delete (DELETE)—thereby evaluating a model’s ability to perform contextually grounded, operationally complete database manipulation.

Across all domains, UPDATE operations constitute 49.64\% of the dataset, demonstrating the benchmark’s emphasis on state-altering reasoning rather than mere data retrieval. In comparison, SELECT, INSERT, and DELETE operations account for 28.93\%, 10.63\%, and 10.80\%, respectively. This distribution highlights our design goal: to assess a model’s capability to handle dynamic, real-world problem-solving, where SQL interactions frequently involve iterative updates, record insertions, and condition-based deletions rather than isolated selection queries.

By jointly modeling multi-turn dialogue, stateful CRUD operations, and domain diversity, this benchmark provides a comprehensive and realistic testbed for evaluating large language models’ proficiency in interactive database intelligence.

\begin{figure}[t]
\centering
\includegraphics[width=0.35\textwidth]{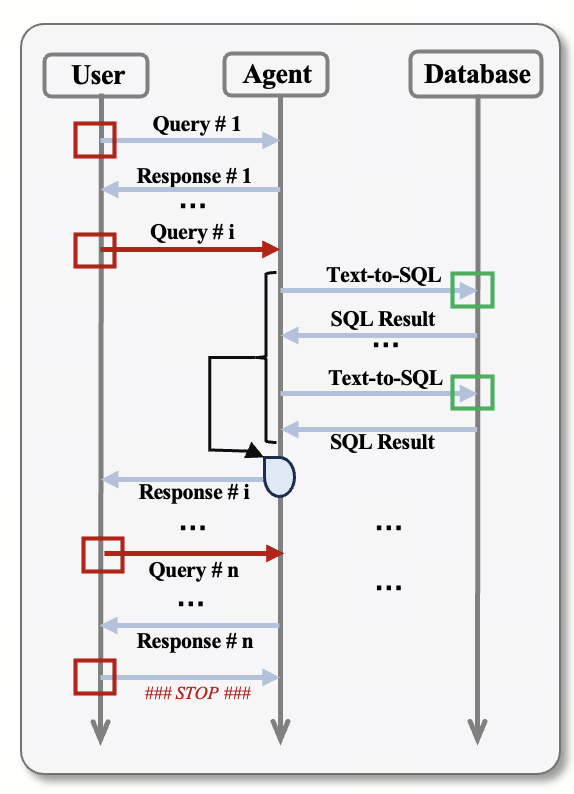}
\caption{Schematic diagram of dynamic multi-turn interactions among the three roles of User, Agent, and Database.}
\label{fig:dynamic_eval}
\end{figure}

\section{Dynamic Interaction}

Our benchmark is designed around a triadic interaction framework (Figure~\ref{fig:dynamic_eval}) involving a simulated user, 
an evaluated model serving as the user-facing agent, and an executable database environment.

\paragraph{\textbf{Interaction Roles}} 
\begin{itemize}
    \item \textbf{Simulated User.} The user is simulated by Qwen2.5-72B-Instruct, where the system message is initialized with the task-specific instruction. An example of instruction is shown in Figure~\ref{fig:user_system_prompt}. 

    \begin{figure}[htbp]
        \centering
        \includegraphics[width=0.45\textwidth]{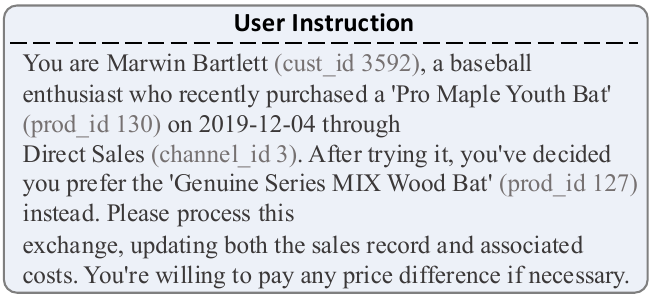}
        \caption{User model system prompt}
        \label{fig:user_system_prompt}
    \end{figure}
    
    \item \textbf{Evaluated agent.} The agent under evaluation is provided with database schema information (DDL) to support query generation. 
    
    \item \textbf{Database Environment.} The database is implemented in SQLite, offering a faithful execution environment that enhances the credibility of the benchmark. In total, the benchmark comprises 1,072 instances spanning 13 sub-tasks, each associated with an independent SQLite database.
\end{itemize}


\paragraph{\textbf{Interaction Logic}}
In the first turn, the user initiates the interaction by issuing a request to the agent according to the given system instruction. During subsequent turns, the user dynamically adjusts its responses based on the agent’s outputs, with the overarching goal of fulfilling the original instruction. The agent, in turn, can exhibit three types of behaviors: (i) interact with the user by replying or requesting additional information; (ii) interact with the database by generating and executing SQL queries; and (iii) perform internal reasoning to refine its interaction strategy. The dialogue terminates when the user outputs \textit{$\#\#\#STOP\#\#\#$} or when the number of interaction turns exceeds the predefined limit~$\eta$.

\begin{figure*}[!t]
  \centering
  \begin{subfigure}{0.24\textwidth}
    \centering
    \includegraphics[width=\linewidth]{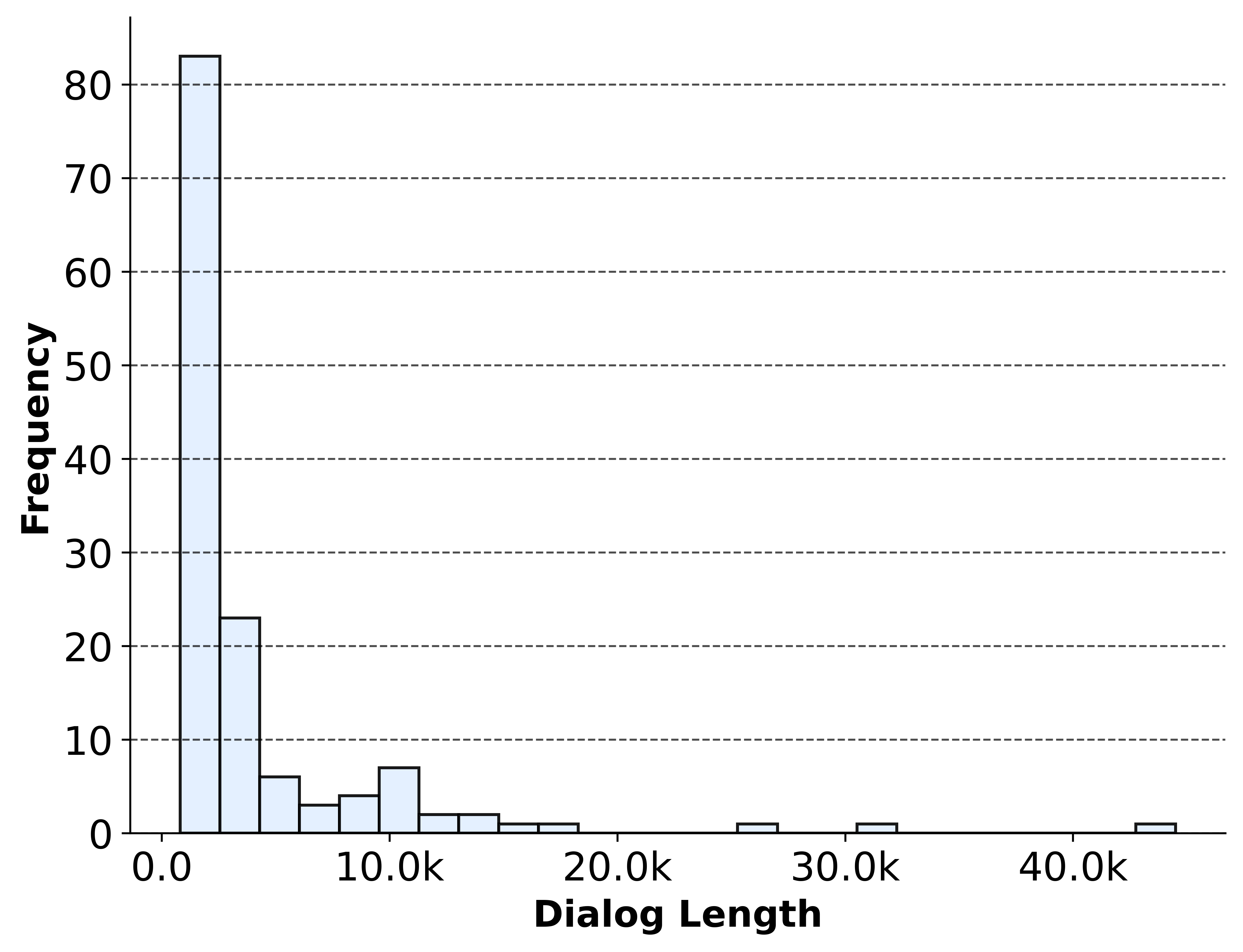}
    \caption{DeepSeek-V3}
    \label{fig:dialog_sub_car_ds}
  \end{subfigure}\hfill
  \begin{subfigure}{0.24\textwidth}
    \centering
    \includegraphics[width=\linewidth]{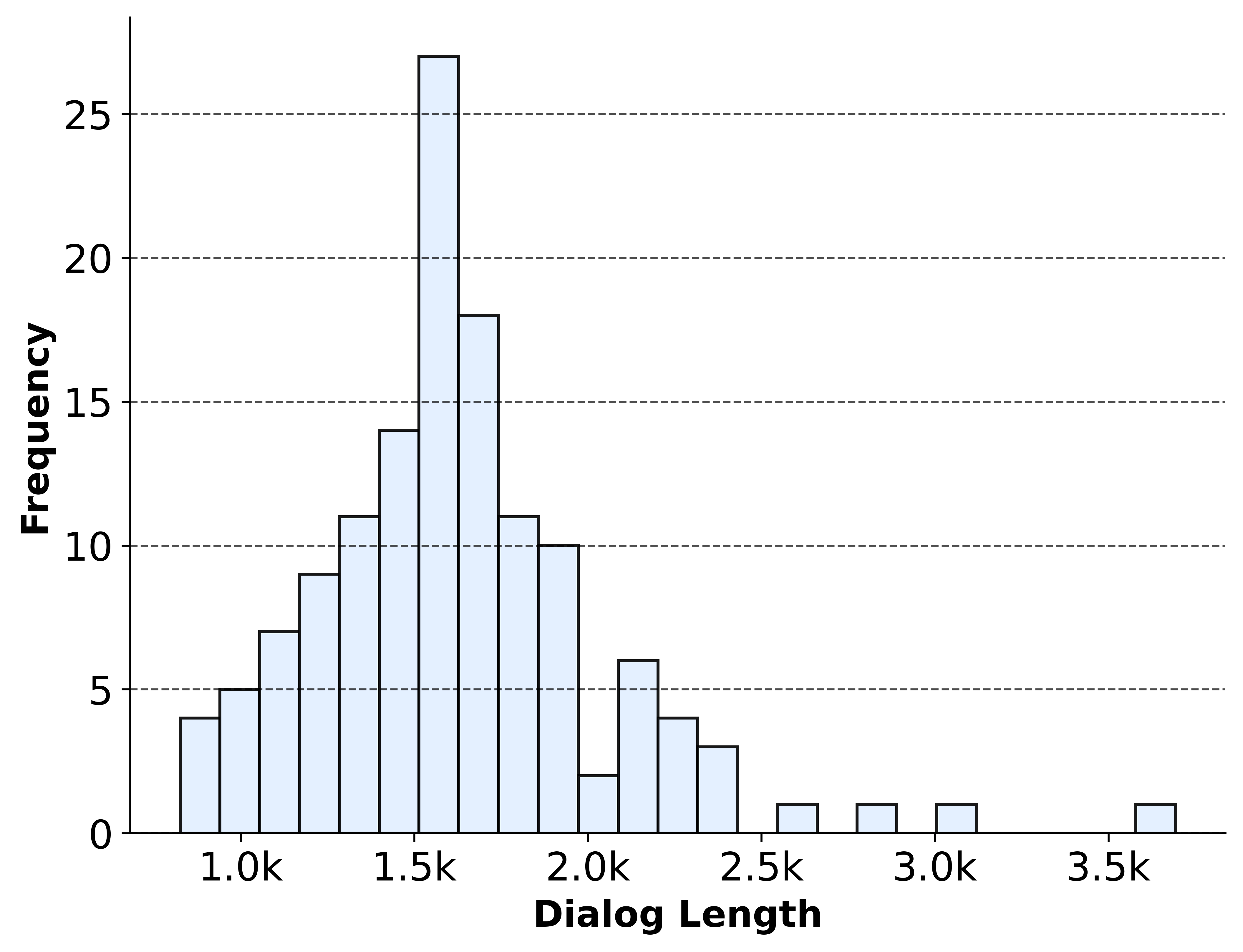}
    \caption{Llama-3.1-70B-Instruct}
    \label{fig:dialog_sub_car_llama}
  \end{subfigure}\hfill
  \begin{subfigure}{0.24\textwidth}
    \centering
    \includegraphics[width=\linewidth]{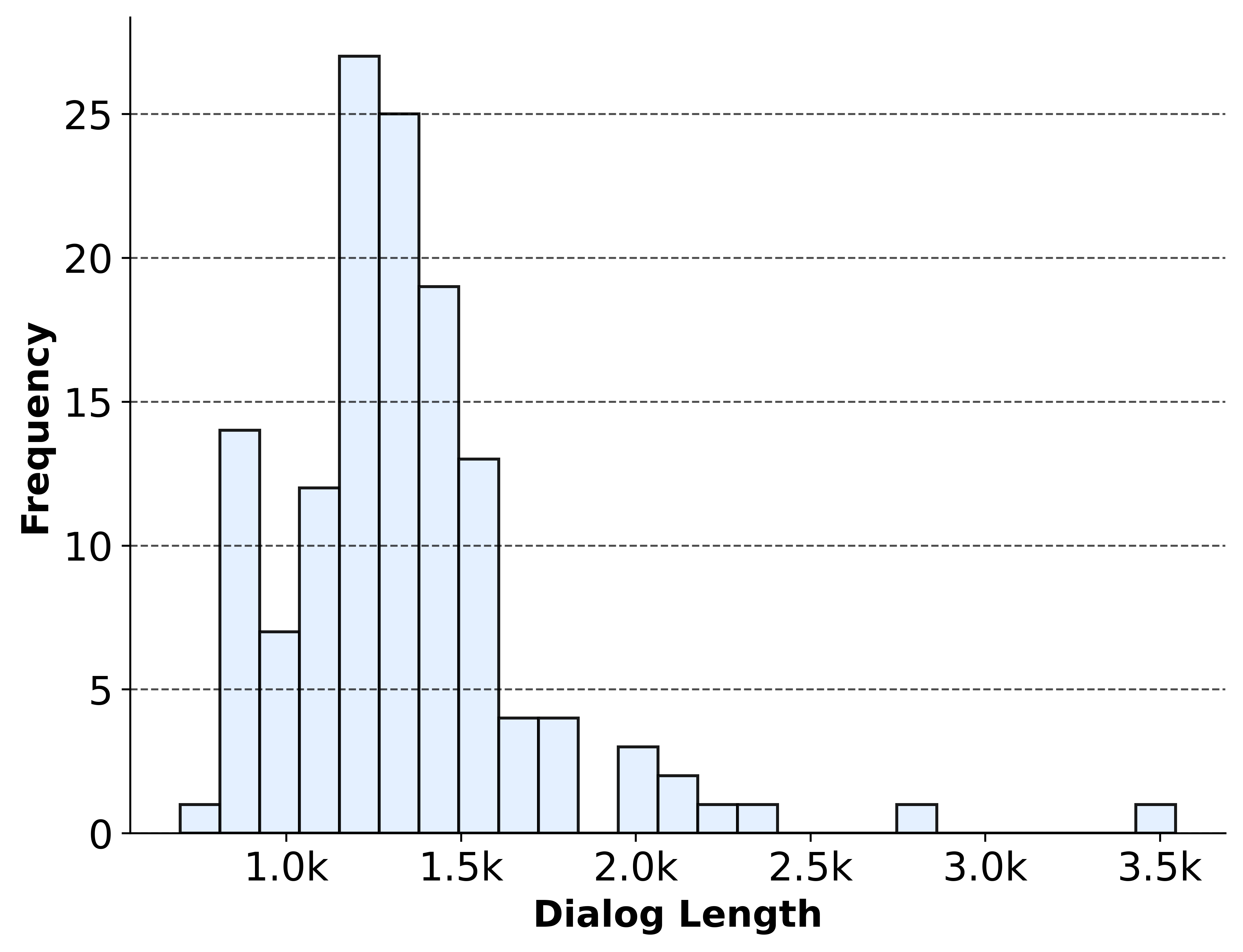}
    \caption{GPT-4o}
    \label{fig:dialog_sub_car_gpt}
  \end{subfigure}\hfill
  \begin{subfigure}{0.24\textwidth}
    \centering
    \includegraphics[width=\linewidth]{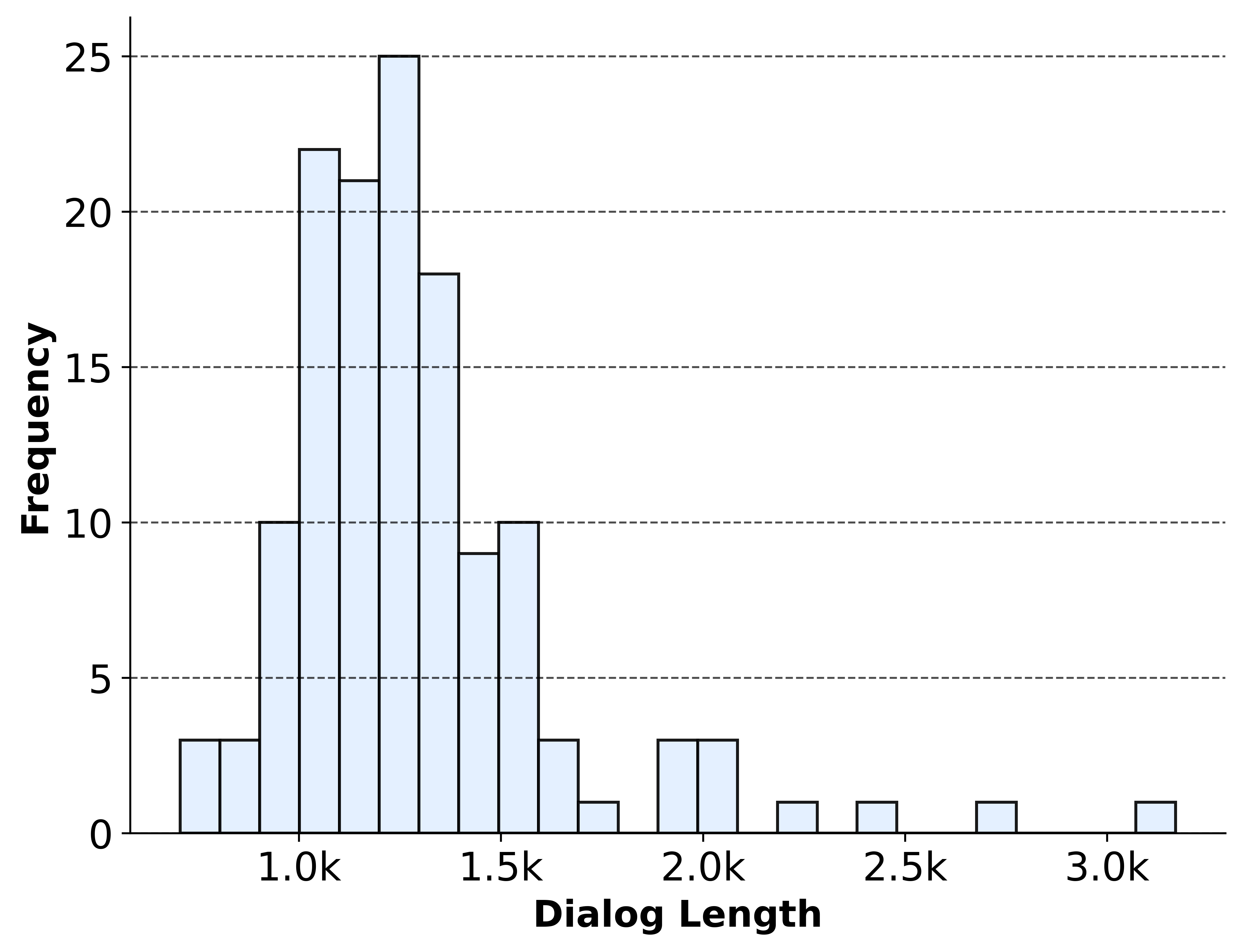}
    \caption{Qwen2.5-72B-Instruct}
    \label{fig:dialog_sub_car_qwen}
  \end{subfigure}

  \vspace{1.5em} 

  \begin{subfigure}{0.24\textwidth}
    \centering
    \includegraphics[width=\linewidth]{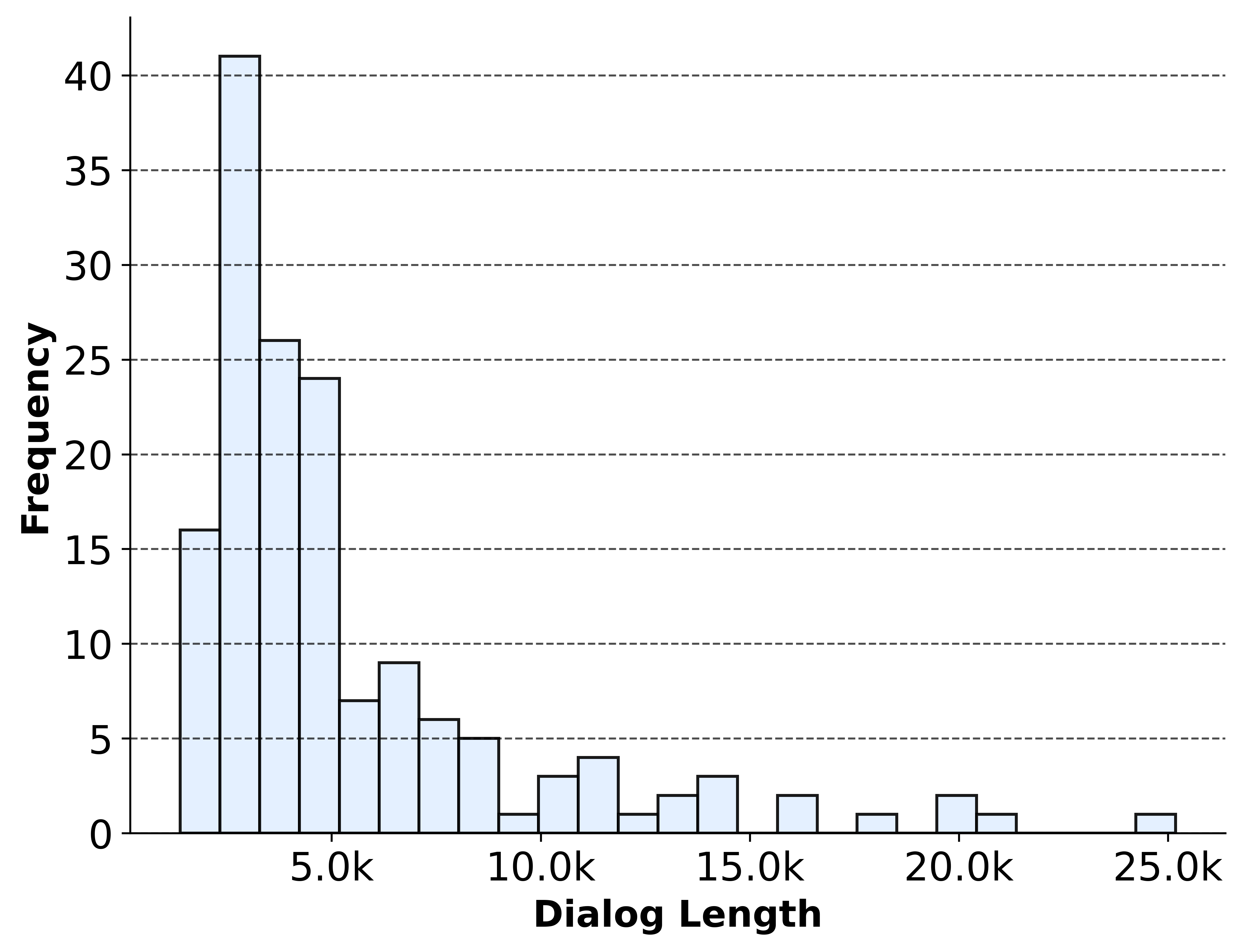}
    \caption{DeepSeek-V3}
    \label{fig:dialog_sub_retail_world_ds}
  \end{subfigure}\hfill
  \begin{subfigure}{0.24\textwidth}
    \centering
    \includegraphics[width=\linewidth]{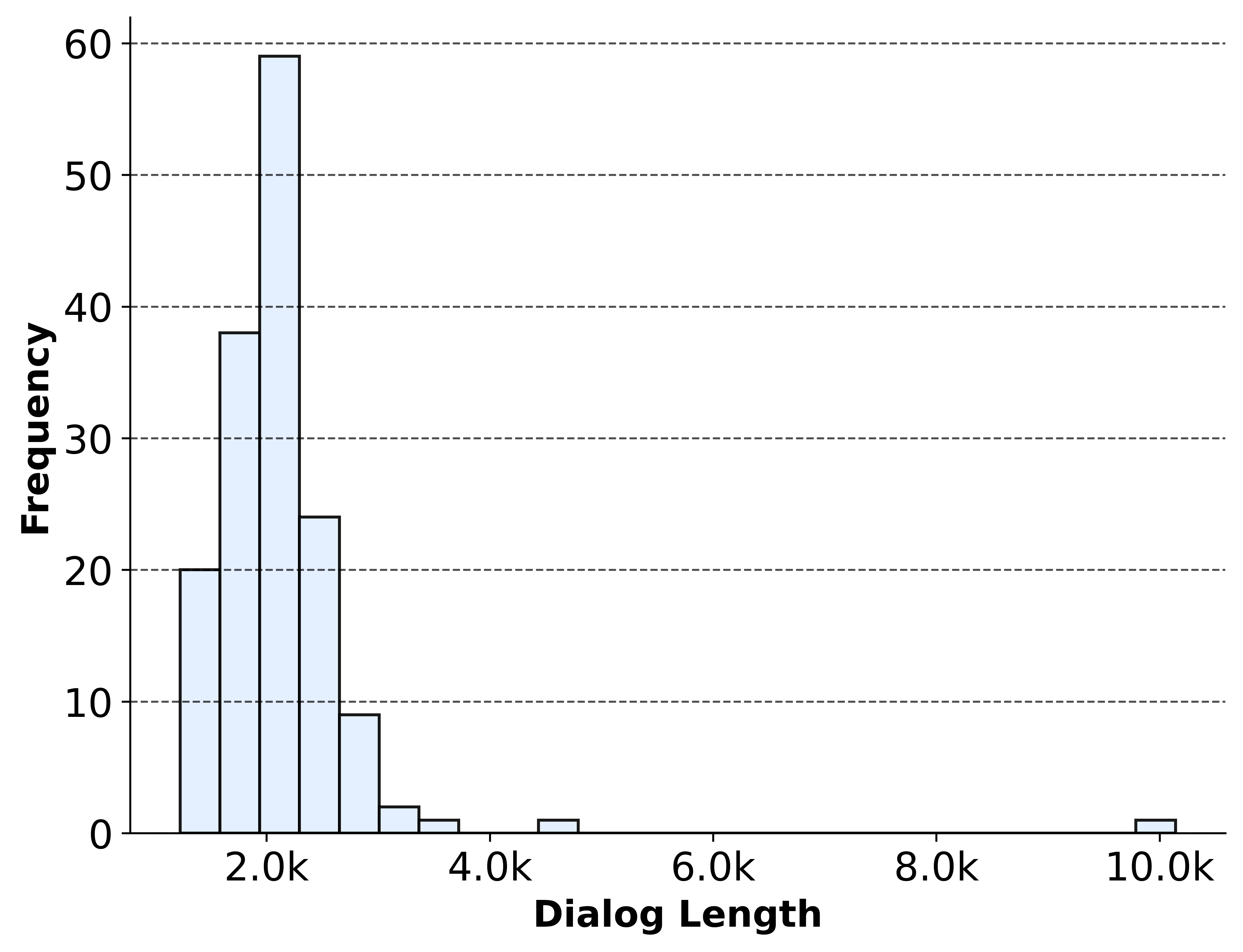}
    \caption{Llama-3.1-70B-Instruct}
    \label{fig:dialog_sub_retail_world_llama}
  \end{subfigure}\hfill
  \begin{subfigure}{0.24\textwidth}
    \centering
    \includegraphics[width=\linewidth]{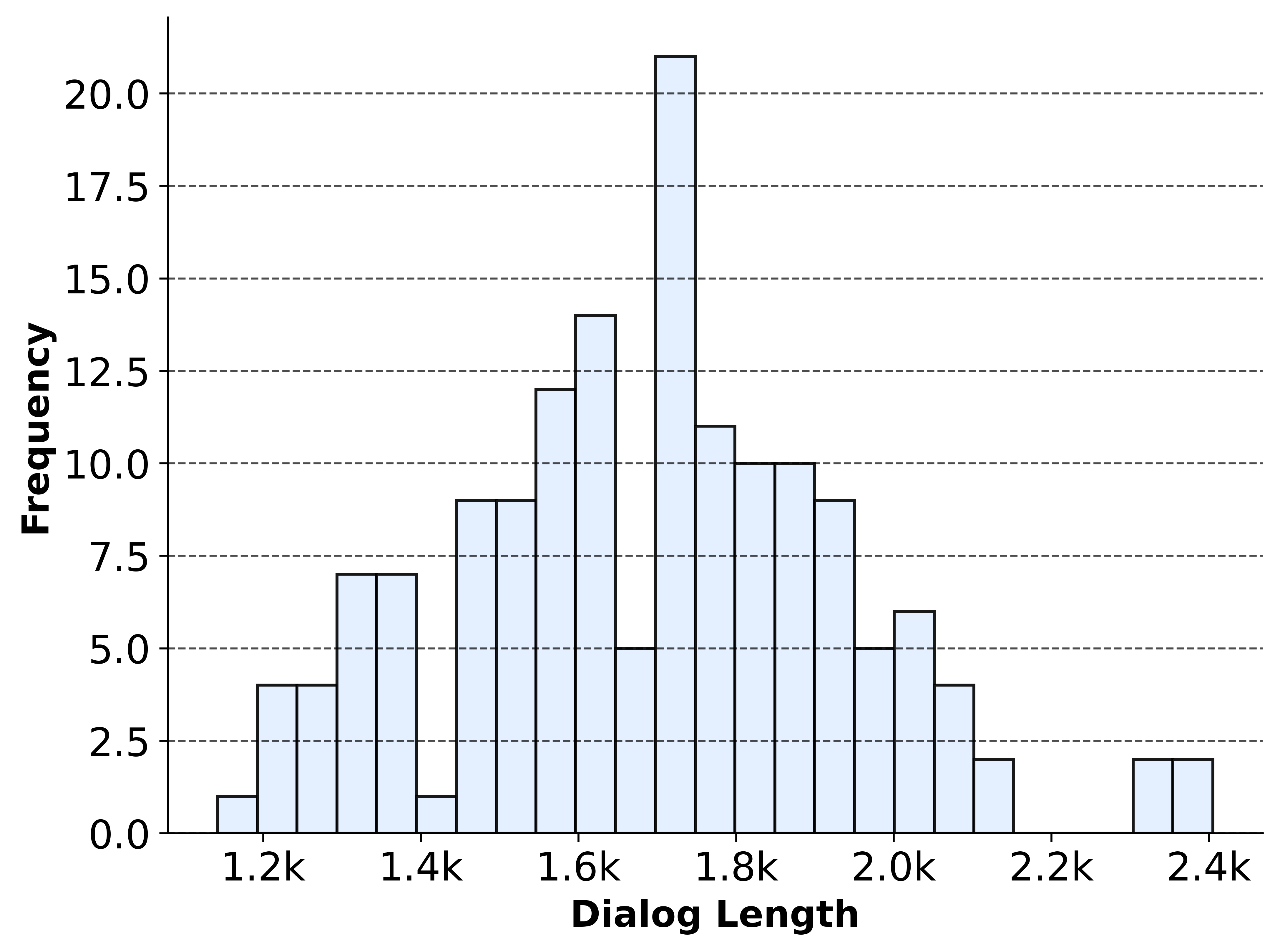}
    \caption{GPT-4o}
    \label{fig:dialog_sub_retail_world_gpt}
  \end{subfigure}\hfill
  \begin{subfigure}{0.24\textwidth}
    \centering
    \includegraphics[width=\linewidth]{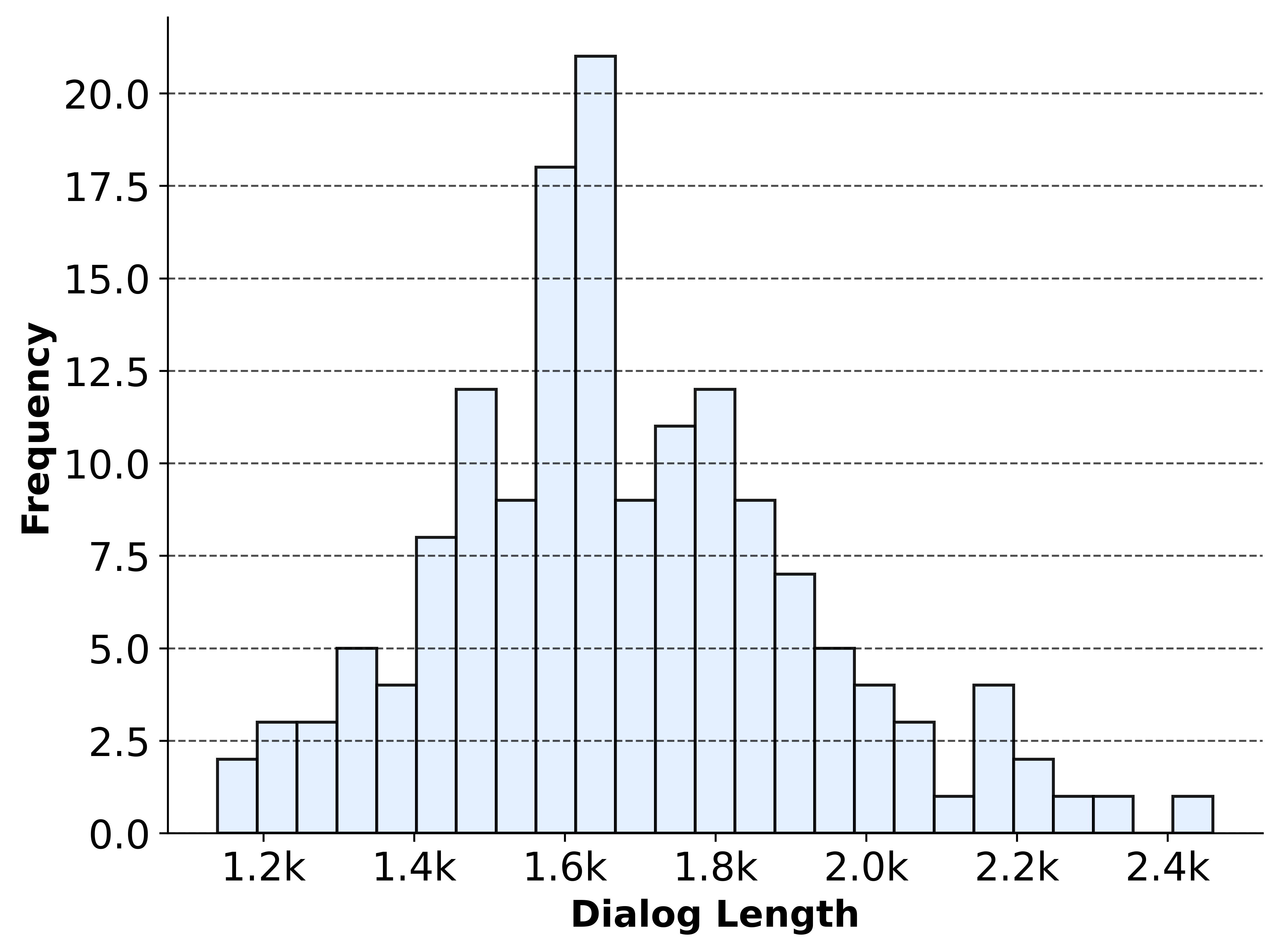}
    \caption{Qwen2.5-72B-Instruct}
    \label{fig:dialog_sub_retail_world_qwen}
  \end{subfigure}

  \caption{The number of dialogue turns of different models on the databases \textit{car} (top row) and \textit{retail\_world} (bottom row).}
  \label{fig:dialogue_length_combined}
\end{figure*}

\section{Experiment}
Building on the tasks constructed in the previous section, we conducted dynamic multi-turn dialogue evaluations across a diverse set of models, encompassing both open-source and proprietary systems. In this section, we aim to address the following key research questions:

\begin{itemize}

\item \textbf{Q1: What overall performance patterns emerge across models of different scales on DySQL-Bench?} (\ref{sec:results}) \\
\item \textbf{Q2: How stable are models under repeated multi-turn SQL interactions?} (\ref{sec:consistency}) \\ 
\item \textbf{Q3: What specific form does hallucination take in dynamic multi-turn SQL interactions, and how does it manifest as extrinsic hallucination?} (\ref{sec:hallucination}) \\
\item \textbf{Q4: What patterns emerge in dialogue length and erroneous SQL invocations across models?} (\ref{sec:dialogue turns and error sql turns}) \\ 
\item \textbf{Q5: What characteristic failure modes emerge in multi-turn SQL dialogues?} (\ref{sec:case study}) \\
\item \textbf{Q6: How effective is few-shot prompting in enhancing SQL models’ reasoning and generalization capabilities under dynamic Text-to-SQL evaluation?} (\ref{sec:few-shot study})\\

\end{itemize}

\subsection{Experimental Setup}\label{sec:setting}
\subsubsection{Hyperparameters} We conducted a systematic evaluation of a wide range of open-source and closed-source models, including GPT-4o~\cite{hurst2024gpt}, DeepSeek-V3~\cite{liu2024deepseek}, Qwen2.5-Max, Qwen2.5-72B-Instruct~\cite{qwen2.5}, Llama-3.1-70B-Instruct~\cite{vavekanand2024llama}, OmniSQL-32B~\cite{li2025omnisql}, Qwen3-32B~\cite{yang2025qwen3}, Qwen3-Coder-30B-A3B-Instruct~\cite{yang2025qwen3} and Gemini-2.5-Flash~\cite{comanici2025gemini}. We set the maximum number of dialogue turns~$\eta$ to 30, and fixed the \textit{temperature} to 0.6, \textit{top\_p} to 0.95, and \textit{top\_k} to 20 for all tested models. Qwen2.5-72B-Instruct was employed as the model simulating the user.
\subsubsection{Metric} We adopt the \textbf{Pass\^{}k} metric proposed in \cite{yao2024tau}, this metric is defined as the probability that \textit{k} i.i.d. solution samples for a given task are all correct, averaged across a distribution of tasks. Formally, it is calculated as follows:

\begin{equation}
\text{Pass\^{}k} = \mathbb{E}_{\text{task}} \left[ \binom{c}{k} / \binom{n}{k} \right]
\label{eq:pass_hat_k}
\end{equation}

Here, for a single task where the model is run for $n$ trials to generate $n$ solutions, $c$ of which are successful, the fraction $\mathbb{E}_{\text{task}}[\binom{c}{k}/\binom{n}{k}]$ represents the probability that a randomly chosen subset of $k$ solutions are all correct. 
In this work, we focus on three specific instances of the metric, namely \textbf{Pass\^{}1}, \textbf{Pass\^{}3}, and \textbf{Pass\^{}5}, which serve as our primary evaluation criteria. The expectation $\mathbb{E}_{\text{task}}[\cdot]$ then averages this probability over all tasks in the evaluation set.



\subsubsection{Implementation Details}
For models with up to 70B parameters, all experiments are conducted on a single server equipped with eight NVIDIA H200 GPUs (each with 140 GB of memory). Using SGLang~\cite{zheng2024sglang}, we deploy the Qwen2.5-72B-Instruct model as the simulated user on four GPUs, while the remaining four GPUs are used to host the agent model under evaluation, also deployed via SGLang. For closed-source models and DeepSeek-r1, we conduct the evaluation through remote API calls.

\subsection{Main Results}\label{sec:results}

As shown in Table~\ref{tab:main-results-all}, model performance on DySQL-Bench generally improves with scale when the parameter size is below approximately 70B, reflecting enhanced reasoning and SQL synthesis capabilities in this range. Beyond this threshold, the performance gains diminish, indicating that model scaling alone is insufficient to ensure further improvements and that closed-source systems still have substantial headroom for optimization on our benchmark.
In the Pass\^{}1 evaluation, Qwen2.5-Max achieves the strongest overall performance, reaching state-of-the-art results on 7 out of 13 databases. Qwen2.5-72B-Instruct also demonstrates competitive results, surpassing Qwen2.5-Max in specific domains such as IH, BO, CH, RE and RW, which highlights its strong effectiveness despite a smaller scale. 
Overall, these findings confirm that scaling up to roughly 70B parameters provides an effective balance between reasoning ability and model size, while further parameter expansion offers diminishing returns, emphasizing the need for targeted optimization and stability improvements beyond mere model size.

\begin{figure*}[!t] 
  \centering
  \begin{subfigure}{0.33\textwidth}
    \centering
    \includegraphics[width=\linewidth]{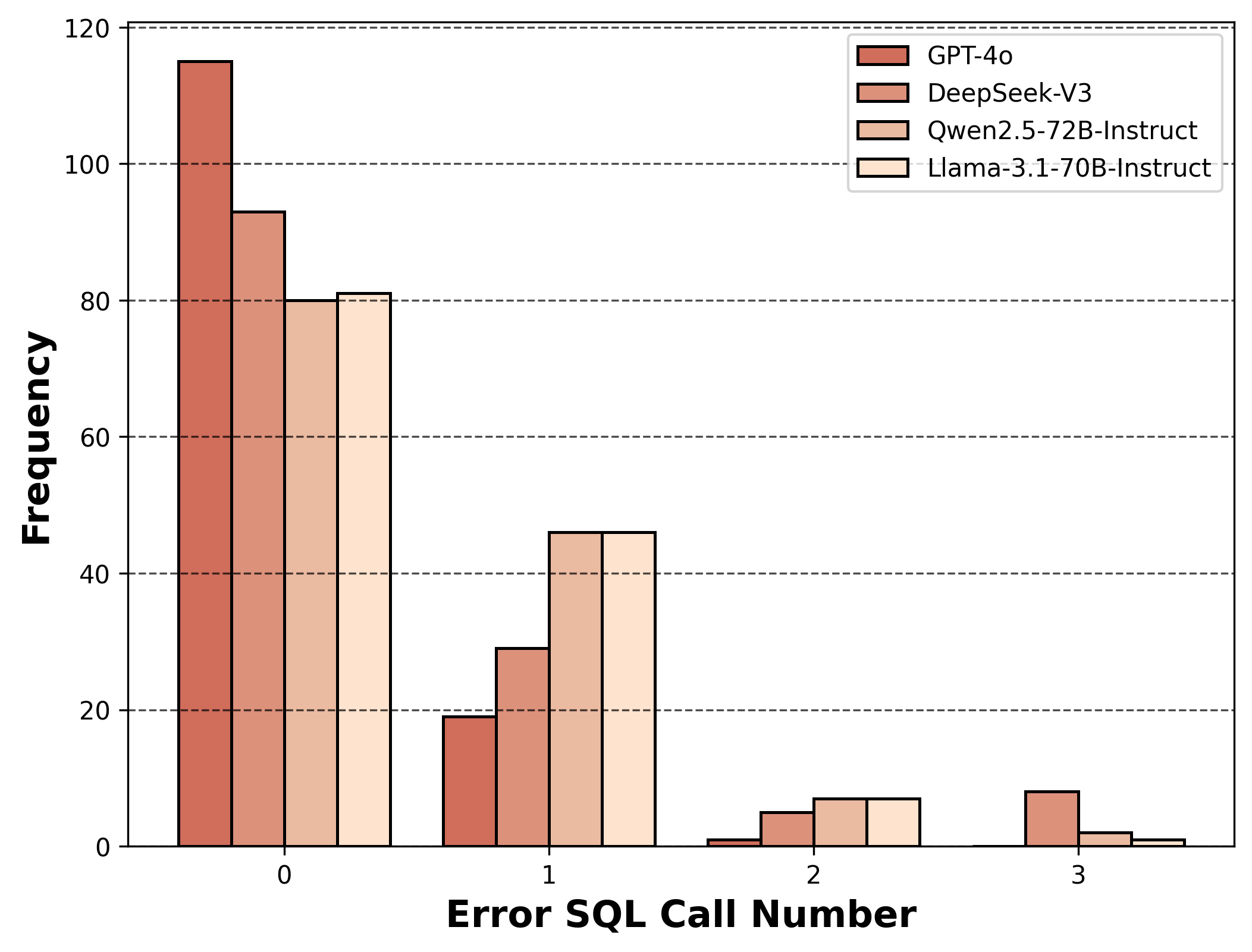}
    \caption{\textit{car}.}
    \label{fig:error_number_car}
  \end{subfigure}\hfill
  \begin{subfigure}{0.33\textwidth}
    \centering
    \includegraphics[width=\linewidth]{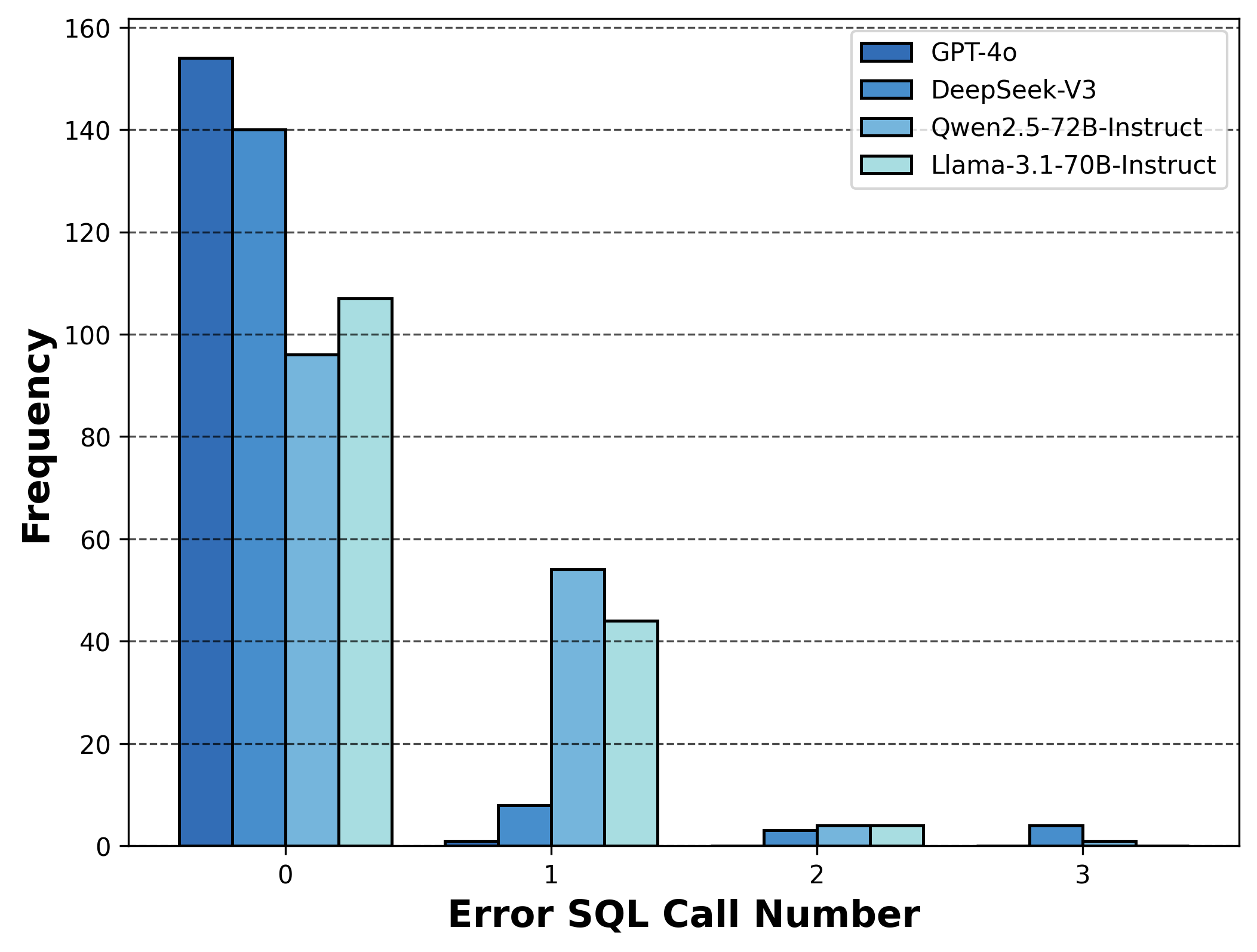}
    \caption{\textit{retail\_world}.}
    \label{fig:error_number_retail_world}
  \end{subfigure}\hfill
  \begin{subfigure}{0.33\textwidth}
    \centering
    \includegraphics[width=\linewidth]{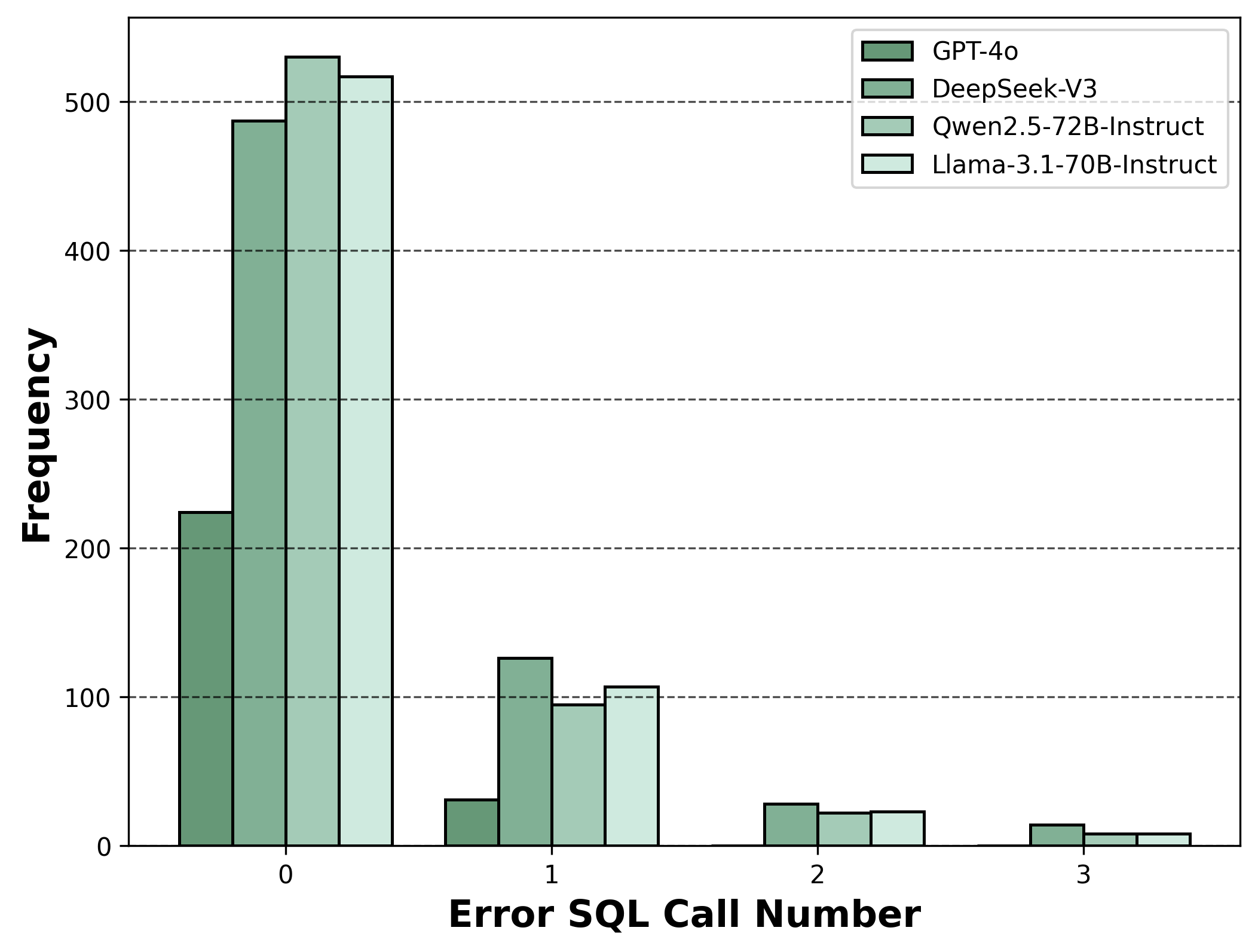}
    \caption{\textit{entertainment}.}
    \label{fig:error_number_entertainment}
  \end{subfigure}\hfill

 \caption{Comparison of the number of dialogue turns with failed SQL invocations for GPT-4o, DeepSeek-V3, Qwen2.5-72B-Instruct, and Llama-3.1-70B-Instruct on the \textit{entertainment}, \textit{car}, and \textit{retail\_world} databases.}
 \label{fig:error_number}
\end{figure*}

\begin{figure}[htbp]
    \centering
    \includegraphics[width=0.45\textwidth]{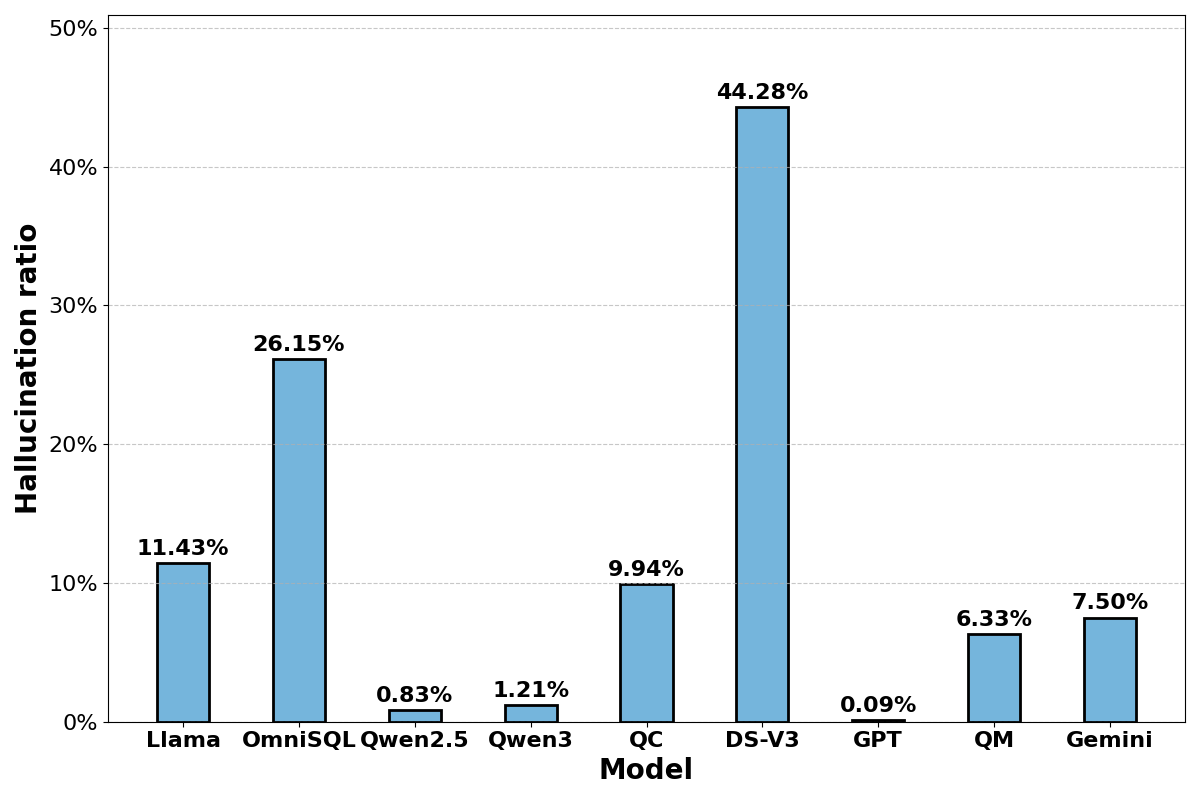}
    \caption{Comparison of hallucination rates across different models. Abbreviation: Llama = Llama3.1-70B-Instruct, OmniSQL = OmniSQL-32B, Qwen2.5-72B-Instruct = Qwen2.5, Qwen3 = Qwen3-32B, QC = Qwen3-Coder-30B-A3B-Instruct, DS-V3 = DeepSeek-V3, GPT = GPT-4o, QM = Qwen2.5-Max, Gemini = Gemini-2.5-Flash.}
    \label{fig:hallu-statistics}
\end{figure}

\begin{figure}[htbp]
    \centering
    \includegraphics[width=0.45\textwidth]{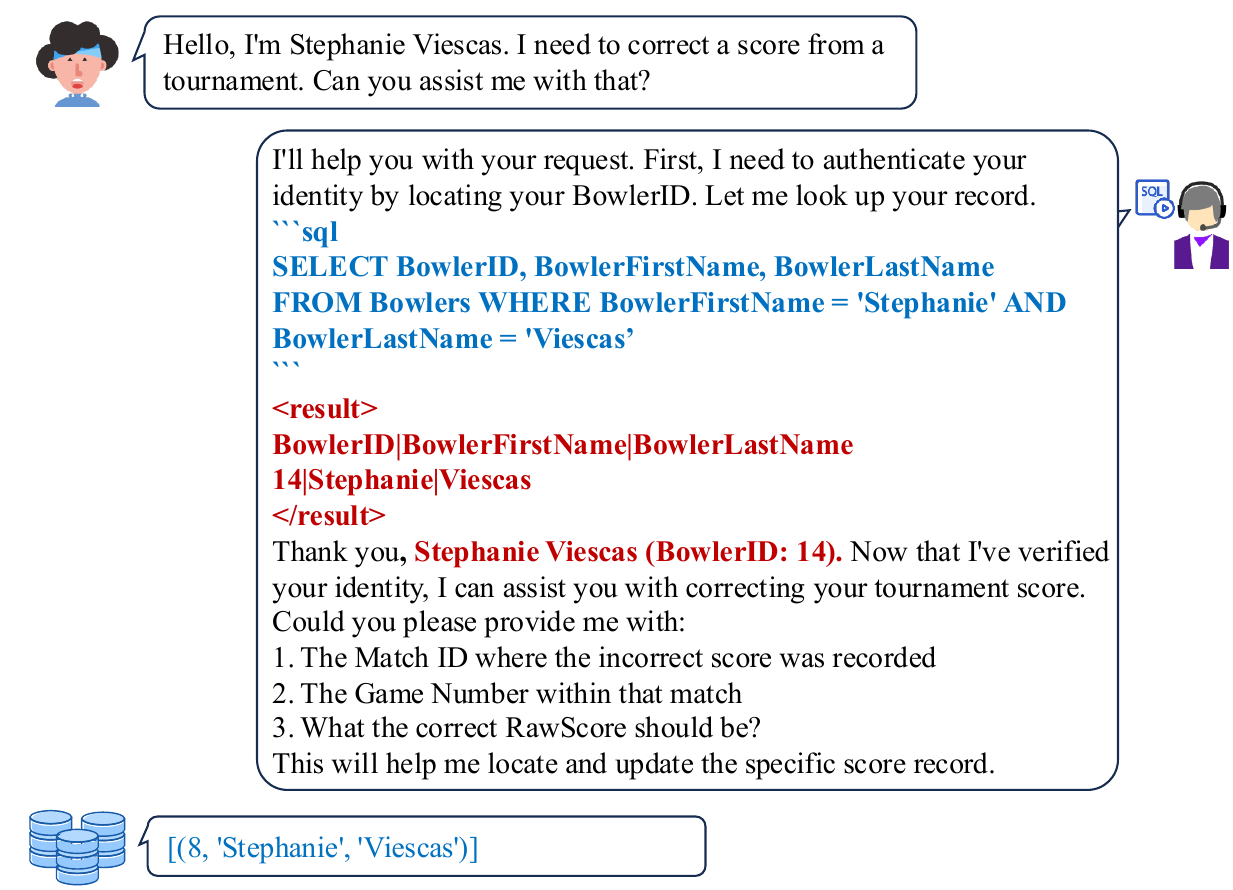}
    \caption{An example of a hallucination produced by the DeepSeek-V3 model. In this example, the user did not provide their own BowlerID. After generating the SQL code block, the model fabricated a BowlerID of 14 and claimed that the user’s identity was successfully verified. However, according to the actual query results, the user’s true BowlerID is 8. This demonstrates that the model produced a factually incorrect output due to hallucination.}
    \label{fig:hallu-example}
\end{figure}

\subsection{Consistency and Stability Analysis Across Multiple Trials} \label{sec:consistency}
As shown in Table~\ref{tab:main-results-all}, and further illustrated in Tables~\ref{tab:main-results-Easy} and~\ref{tab:main-results-hard}, regardless of whether under the \textit{Short} or \textit{Long} complexity, as the number of trial increases, the accuracy of all models drops significantly, indicating that both open-source and proprietary models still have substantial room for improvement in maintaining stable performance during interactions with the execution environment on our benchmark. 
The Pass\^{}k metric plays a crucial role in capturing this phenomenon, as it reflects the model’s ability to consistently and reliably satisfy user intents across repeated multi-turn interactions. Unlike single-pass evaluations, Pass\^{}k directly measures the stability of end-to-end reasoning under stochastic behaviors of both the user simulator and the agent, providing a more faithful assessment of real-world reliability.
Despite the overall decline, the relative ranking among models remains largely consistent across different Pass\^{}k settings. Notably, Qwen2.5-72B-Instruct and Qwen2.5-Max consistently achieve the highest accuracy across most databases, demonstrating robust reasoning stability. In contrast, models such as DeepSeek-V3, Gemini-2.5-flash, Qwen3-32B, and Qwen3-Coder-30B-A3B-Instruct experience rapid performance degradation as the number of trials increases, with some tasks eventually reaching an accuracy of zero. These results highlight the challenge of ensuring response consistency in dynamic multi-turn SQL generation, where models' behavior can vary substantially across repeated executions. In contrast, relying solely on Pass\^{}1 can lead to considerable variance in results, as a single interaction may be influenced by random fluctuations in model behavior or user simulation. Evaluating across multiple trials therefore mitigates such randomness and yields a more stable and trustworthy measurement of performance consistency.


\begin{table*}[htbp]
  \centering
  \caption{Long. Performance comparison of open-source and proprietary models on the Real-World-SQL-Bench benchmark.}
  \resizebox{1.0\linewidth}{!}{
    \begin{tabular}{l|ccc|cccccc|cccc}
    \toprule
    \multirow{2}[4]{*}{\textbf{Model}} & \multicolumn{3}{c|}{\textbf{Sports Domain}} & \multicolumn{6}{c|}{\textbf{Entertainment Domain}} & \multicolumn{4}{c}{\textbf{Business Domain}} \\
\cmidrule{2-14}          & \textbf{ES} & \textbf{IH} & \textbf{BO} & \textbf{EN} & \textbf{MU} & \textbf{LE} & \textbf{CK} & \textbf{CH} & \textbf{PA} & \textbf{CA} & \textbf{HR} & \textbf{RE} & \textbf{RW} \\
    \midrule
    \rowcolor[rgb]{.867, .922, .969}
    \multicolumn{14}{c}{\textit{\textbf{Pass\^{}1}}} \\
    \midrule
    GPT-4o & 59.68 & 33.33 & \underline{71.25} & \underline{64.71} & \underline{64.71} & \underline{38.46} & \textbf{38.10} & 66.67 & \textbf{71.19} & \underline{50.00} & \textbf{68.75} & \textbf{48.86} & \underline{72.22} \\
    DeepSeek-V3 & 34.43 & 16.67 & 26.25 & 48.24 & 29.41 & 30.77 & 23.81 & 33.33 & 22.03 & 25.00 & 43.75 & 17.78 & \underline{72.22} \\
    Gemini2.5-flash & 35.48 & 8.33 & 38.75 & 50.59 & 11.76 & 15.38 & 9.52 & 33.33 & 10.17 & 25.00 & 6.25 & 0.00 & 22.22 \\
    Qwen2.5-Max & \underline{75.41} & 33.33 & \textbf{72.50} & \textbf{68.24} & \underline{64.71} & \underline{38.46} & \textbf{38.10} & \textbf{75.00} & \underline{67.80} & \textbf{75.00} & 56.25 & 37.78 & \textbf{94.44} \\
    Qwen2.5-72B-Instruct & \textbf{75.81} & \textbf{66.67} & 70.00 & 63.53 & \underline{64.71} & \textbf{46.15} & \underline{33.33} & \underline{70.83} & 54.24 & \textbf{75.00} & \underline{62.50} & 42.22 & \textbf{94.44} \\
    Llama3.1-70B-Instruct & 56.45 & 8.33 & 45.00 & 57.65 & 41.18 & \textbf{46.15} & 28.57 & 66.67 & 37.29 & \underline{50.00} & \underline{62.50} & 25.56 & \underline{72.22} \\
    OmniSQL-32B & 43.55 & 33.33 & 28.75 & 43.53 & 29.41 & \textbf{46.15} & 47.62 & 45.83 & 22.03 & \underline{50.00} & 43.75 & 17.78 & 55.56 \\
    Qwen3-32B & 51.61 & \underline{41.67} & 33.75 & 47.06 & \textbf{70.59} & 23.08 & 23.81 & 54.17 & 49.15 & \textbf{75.00} & 50.00 & \underline{43.33} & 38.89 \\
    \midrule
    \rowcolor[rgb]{.867, .922, .969}
    \multicolumn{14}{c}{\textit{\textbf{Pass\^{}3}}} \\
    \midrule
    GPT-4o & 20.97 & 8.33 & 48.75 & 37.65 & \underline{41.18} & 15.38 & 9.52 & 41.67 & \underline{33.90} & \underline{25.00} & 37.50 & \textbf{25.56} & 55.56 \\
    DeepSeek-V3 & 16.13 & 8.33 & 5.00 & 20.00 & 0.00 & 15.38 & 4.76 & 4.17 & 8.47 & 0.00 & 31.25 & 7.78 & 27.78 \\
    Gemini2.5-flash & 16.13 & 0.00 & 15.00 & 5.88 & 0.00 & 0.00 & 4.76 & 0.00 & 3.39 & 0.00 & 6.25 & 0.00 & 5.56 \\
    Qwen2.5-Max & \underline{41.94} & \underline{16.67} & \textbf{52.50} & \underline{40.00} & 23.53 & \underline{23.08} & 9.52 & \underline{54.17} & \textbf{49.15} & \underline{25.00} & \underline{43.75} & 20.00 & \underline{72.22} \\
    Qwen2.5-72B-Instruct & \textbf{45.16} & \textbf{25.00} & \underline{50.00} & \textbf{47.06} & \textbf{52.94} & \textbf{38.46} & \textbf{28.57} & \textbf{58.33} & 28.81 & \textbf{50.00} & \textbf{50.00} & \underline{21.11} & \textbf{94.44} \\
    Llama3.1-70B-Instruct & 17.74 & 0.00 & 18.75 & 32.94 & 11.76 & 15.38 & 9.52 & 25.00 & 15.25 & \underline{25.00} & \textbf{50.00} & 8.89 & 61.11 \\
    OmniSQL-32B & 22.58 & 8.33 & 13.75 & 20.00 & 0.00 & 23.08 & \underline{23.81} & 16.67 & 8.47 & 0.00 & 31.25 & 7.78 & 38.89 \\
    Qwen3-32B & 16.13 & \underline{16.67} & 3.75 & 15.29 & 29.41 & 7.69 & 4.76 & 25.00 & 15.25 & 0.00 & 18.75 & 15.56 & 5.56 \\
    \midrule
    \rowcolor[rgb]{.867, .922, .969}
    \multicolumn{14}{c}{\textit{\textbf{Pass\^{}5}}} \\
    \midrule
    GPT-4o & 9.68 & 0.00 & \textbf{37.50} & \underline{28.24} & \textbf{35.29} & 0.00 & 4.76 & 29.17 & 18.64 & \underline{25.00} & 25.00 & \textbf{15.56} & 55.56 \\
    DeepSeek-V3 & 6.45 & \underline{8.33} & 5.00 & 14.12 & 0.00 & 7.69 & 4.76 & 4.17 & 3.39 & 0.00 & 25.00 & 6.67 & 22.22 \\
    Gemini2.5-flash & 3.23 & 0.00 & 7.50 & 0.00 & 0.00 & 0.00 & 0.00 & 0.00 & 3.39 & 0.00 & 6.25 & 0.00 & 0.00 \\
    Qwen2.5-Max & \underline{29.03} & \underline{8.33} & \underline{35.00} & 24.71 & \underline{17.65} & \underline{15.38} & 4.76 & \underline{37.50} & \textbf{35.59} & \underline{25.00} & \textbf{43.75} & \underline{13.33} & \underline{66.67} \\
    Qwen2.5-72B-Instruct & \textbf{32.26} & \textbf{25.00} & 33.75 & \textbf{40.00} & \textbf{35.29} & \textbf{38.46} & \underline{9.52} & \textbf{50.00} & \underline{22.03} & \textbf{50.00} & \textbf{43.75} & \underline{13.33} & \textbf{77.78} \\
    Llama3.1-70B-Instruct & 11.29 & 0.00 & 12.50 & 24.71 & 5.88 & \underline{15.38} & \underline{9.52} & 20.83 & 8.47 & 0.00 & \textbf{43.75} & 5.56 & 55.56 \\
    OmniSQL-32B & 9.68 & \underline{8.33} & 7.50 & 15.29 & 0.00 & 7.69 & \textbf{14.29} & 4.17 & 5.08 & 0.00 & \underline{31.25} & 5.56 & 38.89 \\
    Qwen3-32B & 8.06 & \underline{8.33} & 2.50 & 8.24 & 11.76 & 7.69 & 4.76 & 20.83 & 8.47 & 0.00 & 6.25 & 8.89 & 5.56 \\
    \bottomrule
    \end{tabular}%
    }
  \label{tab:main-results-hard}%
\end{table*}%

\begin{table*}[t]
\caption{ALL. Performance comparison of open-source and proprietary models on the Real-World-SQL-Bench benchmark.}
\resizebox{1.0\linewidth}{!}{
\label{tab:main-results-all}
\centering
\begin{tabular*}{\textwidth}{@{\extracolsep{\fill}} l| ccc|cccccc|cccc} 
\toprule
    \multirow{2}[4]{*}{\textbf{Model}} & \multicolumn{3}{c|}{\textbf{Sports Domain}} & \multicolumn{6}{c|}{\textbf{Entertainment Domain}} & \multicolumn{4}{c}{\textbf{Business Domain}} \\
\cmidrule{2-14}          & \textbf{ES} & \textbf{IH} & \textbf{BO} & \textbf{EN} & \textbf{MU} & \textbf{LE} & \textbf{CK} & \textbf{CH} & \textbf{PA} & \textbf{CA} & \textbf{HR} & \textbf{RE} & \textbf{RW} \\
\midrule
\rowcolor[rgb]{.867, .922, .969}
\multicolumn{14}{c}{\textit{\textbf{Pass\^{}1}}} \\
\midrule
GPT-4o & 62.79 & \underline{46.43} & 70.27 & 70.23 & 66.67 & 30.19 & 33.33 & \underline{68.18} & \underline{65.71} & 51.85 & 62.50 & 52.20 & 77.42 \\
DeepSeek-V3 & 48.84 & 28.57 & 26.13 & 56.49 & 33.33 & 33.96 & 31.37 & 43.18 & 36.19 & 40.74 & 45.00 & 30.73 & \underline{74.19} \\
Gemini2.5-flash & 48.37 & 14.29 & 41.44 & 51.91 & 14.29 & 15.09 & 21.57 & 34.09 & 13.33 & 22.22 & 15.00 & 0.98 & 35.48 \\
Qwen2.5-Max & \textbf{76.74} & \underline{46.43} & \underline{73.87} & \textbf{74.81} & \underline{71.43} & 41.51 & \textbf{49.02} & \underline{68.18} & \textbf{68.57} & \textbf{59.26} & \textbf{72.50} & \underline{56.59} & \textbf{90.32} \\
Qwen2.5-72B-Instruct & \underline{73.49} & \textbf{67.86} & \textbf{74.77} & \underline{71.76} & 66.67 & 56.60 & \underline{43.14} & \textbf{77.27} & 62.86 & \underline{55.56} & 67.50 & \textbf{57.56} & \textbf{90.32} \\
Llama3.1-70B-Instruct & 56.74 & 35.71 & 46.85 & 65.65 & 42.86 & \underline{60.38} & 29.41 & \underline{68.18} & 46.67 & 48.15 & \underline{70.00} & 40.00 & \underline{74.19} \\
OmniSQL-32B & 57.21 & 32.14 & 37.84 & 47.33 & 28.57 & \textbf{66.04} & 39.22 & 52.27 & 28.57 & 51.85 & 60.00 & 31.22 & 48.39 \\
Qwen3-32B & 55.81 & \underline{46.43} & 30.63 & 51.15 & \textbf{76.19} & 26.42 & 27.45 & 54.55 & 51.43 & \underline{55.56} & 50.00 & 49.27 & 41.94 \\
\midrule
\rowcolor[rgb]{.867, .922, .969}
\multicolumn{14}{c}{\textit{\textbf{Pass\^{}3}}} \\
\midrule
GPT-4o & 30.23 & 25.00 & 45.95 & 46.56 & \underline{47.62} & 7.55 & 11.76 & 38.64 & 36.19 & 22.22 & 40.00 & 27.80 & 64.52 \\
DeepSeek-V3 & 30.70 & 14.29 & 8.11 & 25.95 & 4.76 & 16.98 & 7.84 & 9.09 & 11.43 & 18.52 & 25.00 & 16.10 & 38.71 \\
Gemini2.5-flash & 20.00 & 0.00 & 17.12 & 6.87 & 0.00 & 5.66 & 5.88 & 0.00 & 5.71 & 0.00 & 2.50 & 0.00 & 25.81 \\
Qwen2.5-Max & \underline{52.56} & \underline{32.14} & \underline{53.15} & \underline{47.33} & 38.10 & 24.53 & 17.65 & \underline{54.55} & \textbf{45.71} & \underline{29.63} & \underline{47.50} & \underline{31.71} & \underline{70.97} \\
Qwen2.5-72B-Instruct & \textbf{53.49} & \textbf{35.71} & \textbf{54.05} & \textbf{55.73} & \textbf{57.14} & \textbf{43.40} & \textbf{29.41} & \textbf{61.36} & \underline{40.00} & \textbf{44.44} & \textbf{55.00} & \textbf{38.54} & \textbf{83.87} \\
Llama3.1-70B-Instruct & 26.98 & 21.43 & 22.52 & 45.04 & 14.29 & 28.30 & 7.84 & 38.64 & 23.81 & 22.22 & \underline{47.50} & 17.56 & 67.74 \\
OmniSQL-32B & 36.74 & 21.43 & 19.82 & 25.19 & 0.00 & \underline{33.96} & \underline{21.57} & 25.00 & 13.33 & 14.81 & 40.00 & 15.61 & 32.26 \\
Qwen3-32B & 26.05 & 21.43 & 3.60 & 19.85 & 33.33 & 7.55 & 3.92 & 29.55 & 17.14 & 22.22 & 20.00 & 23.90 & 22.58 \\
\midrule
\rowcolor[rgb]{.867, .922, .969}
\multicolumn{14}{c}{\textit{\textbf{Pass\^{}5}}} \\
\midrule
GPT-4o & 19.53 & 17.86 & 34.23 & 32.82 & \underline{33.33} & 1.89 & 5.88 & 27.27 & 20.95 & 11.11 & 30.00 & 16.59 & 58.06 \\
DeepSeek-V3 & 17.67 & 10.71 & 7.21 & 17.56 & 0.00 & 9.43 & 7.84 & 6.82 & 6.67 & 7.41 & 20.00 & 11.71 & 29.03 \\
Gemini2.5-flash & 6.98 & 0.00 & 8.11 & 0.00 & 0.00 & 3.77 & 1.96 & 0.00 & 4.76 & 0.00 & 2.50 & 0.00 & 19.35 \\
Qwen2.5-Max & \underline{39.53} & \underline{25.00} & \underline{36.04} & \underline{33.59} & 28.57 & 9.43 & 9.80 & \underline{40.91} & \textbf{34.29} & \underline{18.52} & \underline{42.50} & \underline{20.49} & \underline{61.29} \\
Qwen2.5-72B-Instruct & \textbf{44.65} & \textbf{28.57} & \textbf{39.64} & \textbf{48.09} & \textbf{38.10} & \textbf{33.96} & \textbf{19.61} & \textbf{56.82} & \underline{26.67} & \textbf{44.44} & \textbf{52.50} & \textbf{25.85} & \textbf{74.19} \\
Llama3.1-70B-Instruct & 14.88 & 17.86 & 17.12 & 32.82 & 9.52 & \underline{20.75} & 7.84 & 31.82 & 14.29 & 7.41 & 40.00 & 11.22 & \underline{61.29} \\
OmniSQL-32B & 24.65 & 21.43 & 12.61 & 18.32 & 0.00 & 18.87 & \underline{13.73} & 13.64 & 8.57 & 11.11 & 27.50 & 7.80 & 32.26 \\
Qwen3-32B & 14.88 & 3.57 & 2.70 & 12.21 & 19.05 & 5.66 & 3.92 & 18.18 & 9.52 & 7.41 & 7.50 & 13.66 & 22.58 \\
\bottomrule
\end{tabular*}
}
\end{table*}

\subsection{The issue of hallucination in multi-turn dialogue remains a critical challenge for large language models.} \label{sec:hallucination}
During dynamic multi-turn interactions between the agent model and the user model, we found that the agent often exhibits hallucination after generating an SQL query, namely by fabricating the query results on its own. According to the taxonomy of hallucinations defined in prior work~\cite{bang2025hallulens}, this phenomenon falls under \textit{Extrinsic Hallucination}, as it involves generations that are inconsistent with the training data and cannot be verified by the given context. Such hallucinations typically arise when models attempt to fill knowledge gaps or produce unsupported content beyond the scope of the provided input, rather than misinterpreting the input itself. We identify hallucinations by checking whether each model’s dialogue trajectory output includes an SQL code block that is immediately followed by the special token \textit{\textless result\textgreater}, which is used in the agent model’s system prompt to denote the SQL execution result returned from the environment. As illustrated in Figure~\ref{fig:hallu-statistics}, DeepSeek-V3 and OmniSQL-32B demonstrate the highest hallucination rates, at 44.28\% and 26.15\%, respectively. The hallucination rates of Llama3.1-70B-Instruct, Qwen3-Coder-30B-A3B-Instruct, Qwen2.5-Max, and Gemini-2.5-Flash are approximately 10\%, whereas GPT-4o yields the lowest hallucination rate of 0.09\%. We hypothesize that the high hallucination rates observed in DeepSeek-V3 and OmniSQL-32B stem from the models’ tendency to learn step-by-step problem-solving patterns during post-training. In this process, the models are trained to generate solutions in a procedural manner, often providing intermediate results at each step. Consequently, they internalize this paradigm. However, in the SQL-calling scenario, the execution results are produced externally by the environment rather than inferred by the model itself. When the model continues to follow its learned reasoning paradigm in this context, it tends to fabricate results, leading to hallucinations. We present an example in Figure~\ref{fig:hallu-example}. After receiving the user’s instruction, the model first performs identity verification. However, after generating a query to retrieve the user’s ID (\textit{BowlerID}), it fabricates an incorrect result (\textit{BowlerID = 14}), whereas the actual query output should be \textit{BowlerID = 8}. This behavior demonstrates that the model produced a factually incorrect output due to hallucination.

\subsection{Analysis of Dialogue Turns and Erroneous SQL Invocation Turns} \label{sec:dialogue turns and error sql turns}
We analyze how different models perform when SQL execution fails across databases. 
For clarity and conciseness, we select three representative databases, \textit{entertainment}, \textit{retail\_world} and \textit{car}, and use \textbf{GPT-4o}, \textbf{DeepSeek-V3}, \textbf{Qwen2.5-72B-Instruct}, and \textbf{Llama-3.1-70B-Instruct} as examples to illustrate their behaviors on these databases. As shown in Figure~\ref{fig:dialogue_length_combined}, DeepSeek-V3 exhibits a pronounced long-tailed distribution. Its central 60\% of dialogues span a wide range—approximately 2.7k–7.1k tokens on \textit{retail\_world} and 1.5k–4.9k tokens on \textit{car}—while several extreme cases exceed 20k tokens. We attribute this phenomenon to the use of GRPO algorithm~\cite{shao2024deepseekmath} during post-training, which likely encourages the model to generate longer and more exploratory responses, thereby substantially increasing dialogue length. In contrast, GPT-4o and Qwen2.5-72B-Instruct display compact, single-peaked distributions centered around 1.4k–1.9k tokens, reflecting stable clarify–execute behavior and strong schema grounding. Llama-3.1-70B-Instruct follows a similar pattern but with dialogue lengths typically around 1.3k–2.4k tokens. 
As shown in Figure~\ref{fig:error_number}, we observe that GPT-4o attains the highest fraction of zero-error turns on both \textit{car} and \textit{retail\_world}; non-zero errors are rare, with only a small spike at three errors on \textit{car}. Qwen2.5-72B-Instruct and Llama-3.1-70B-Instruct also concentrate at zero but exhibit a more visible single-error bar, consistent with minor, recoverable mismatches. DeepSeek-V3 likewise has many zero-error turns and few multi-error cases, yet its interactions are markedly longer. Compared with Qwen2.5-72B-Instruct, both DeepSeek-V3 and GPT-4o produce SQL invocations that fail less often due to syntax errors. However, their end-to-end task accuracy remains noticeably lower than Qwen2.5’s(see Table~\ref{tab:main-results-all}). This gap suggests that on our benchmark, intent understanding, multi-turn planning, and schema-aware SQL synthesis—rather than surface-level SQL correctness—are the primary bottlenecks for DeepSeek-V3 and GPT-4o, indicating room for improvement in generating higher-quality, goal-satisfying SQL over extended dialogues.
Extending to the \textit{entertainment} domain, we observe that GPT-4o’s proportion of turns with zero SQL invocation errors is significantly lower than that of other models, while its proportions for one, two, and three errors remain comparable. Nevertheless, it maintains near-SOTA results on Pass\^{}1, achieving 80.43\% accuracy on \textit{Short} and 64.71\% on \textit{Long} samples (see Table~\ref{tab:main-results-Easy} and ~\ref{tab:main-results-hard}). This finding suggests that when the task inherently requires a larger number of SQL invocations, GPT-4o is able to balance invocation frequency and overall task performance more effectively, avoiding unnecessary SQL calls while maintaining strong end-to-end accuracy.

\subsection{Case Study} \label{sec:case study}
To better illustrate the behavior of different models in real-world multi-turn SQL interactions, we conduct a case study based on representative dialogue examples from our benchmark in this section. This case shows that the model’s strict adherence to predefined verification rules prevented it from utilizing reasonable contextual cues provided by the user, leading to premature task termination.

\begin{figure}
\includegraphics[width=0.48\textwidth]{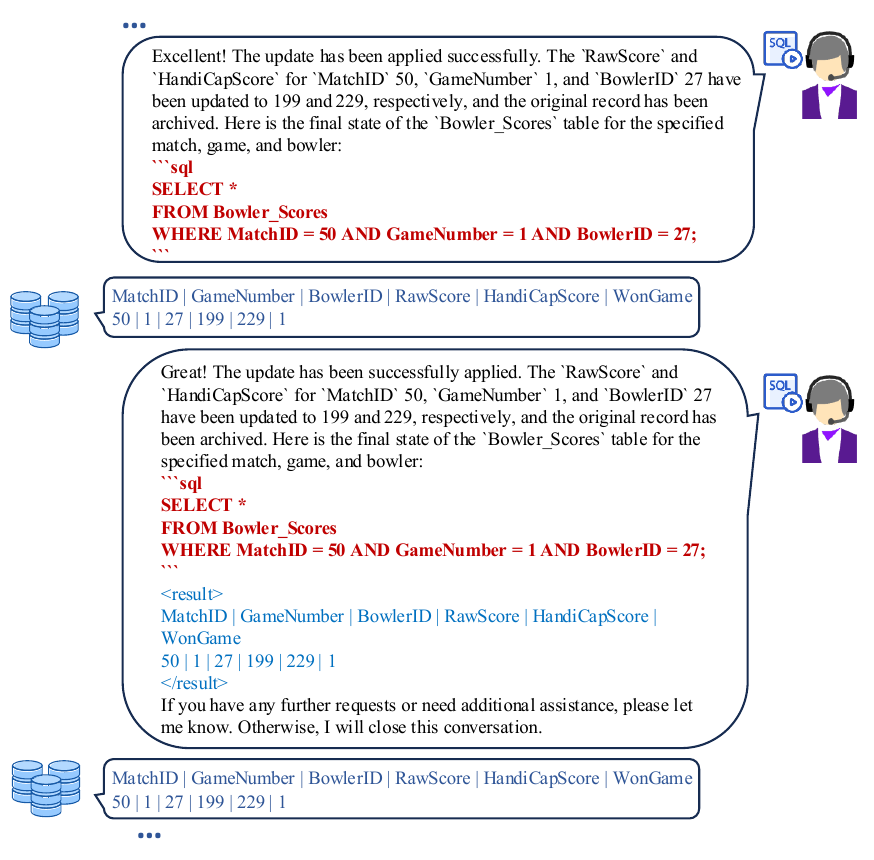}
\caption{Repetition Loop after SQL Execution.}
\label{fig:case_2}
\end{figure}

\paragraph{Repetition Loop after SQL Execution} In Figure~\ref{fig:case_2}, the agent executed the requested SQL correctly yet then entered a repetition loop, repeatedly outputting the same segment until the context buffer was exhausted. This loop is not due to SQL syntax or schema mismatch, but rather a generation-control failure: the model fails to transition from “SQL executed → next step” to “continue dialogue / end” and instead continues repeating. Because each repeated turn contributes no new progress but consumes tokens, such behavior significantly inflates dialogue length without improving task completion. We further observe that this repetition issue occurs frequently in the dialogues generated by OmniSQL-32B, suggesting that while the model's post-training has successfully improved raw Text-to-SQL invocation ability, their multi-turn dialogue capability appears to degrade in parallel - in other words, stronger SQL emission power is accompanied by weaker sustained conversational control. This trend underscores the importance of balancing one-shot SQL competence with multi-turn interaction fluency and termination policies in system design.

\begin{figure}
\includegraphics[width=0.48\textwidth]{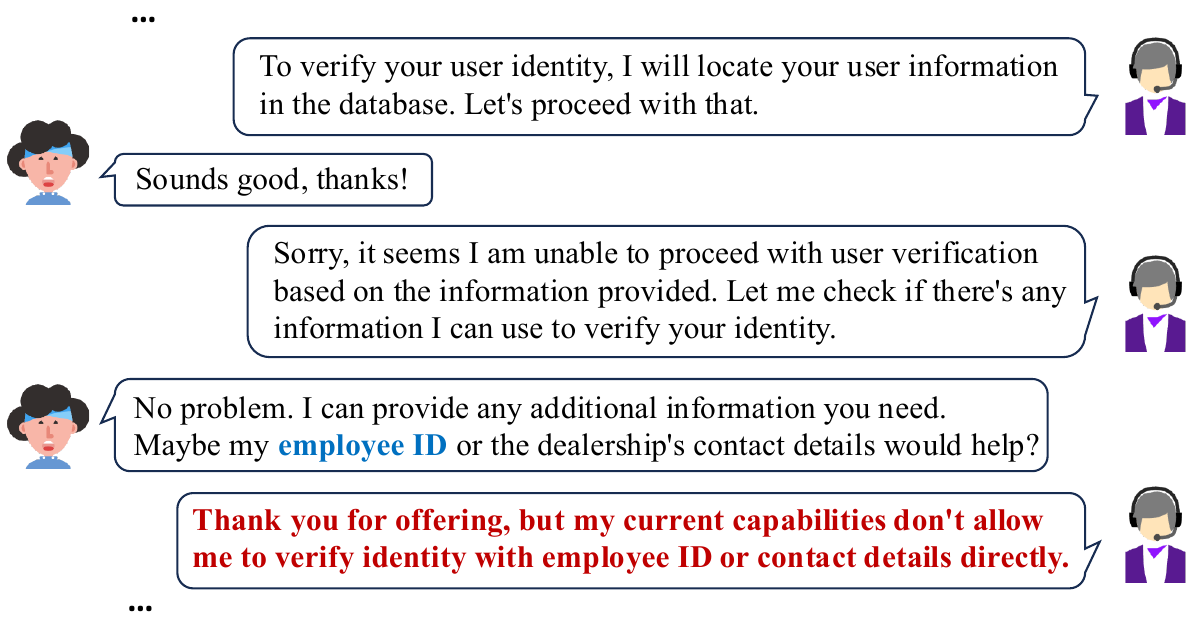}
\caption{Refusal to Use Available User Information for Identity Verification.}
\label{fig:case_1}
\end{figure}

\paragraph{Refusal to Use Available User Information for Identity Verification} As shown in Figure~\ref{fig:case_1}, when the user offered additional identifiers such as an \textit{employee ID} to assist verification, the model explicitly refused to proceed, replying “my current capabilities don’t allow me to verify identity with employee ID or contact details directly.” This indicates that the model failed to leverage available contextual information to complete identity verification. As a result, it terminated the task without executing the intended SQL operations. It highlights the need for more pragmatic interaction handling mechanisms that allow models to proceed with the task when sufficient user intent and information are available.

\subsection{Few-shot Study} \label{sec:few-shot study}

In the above experiments, all results are obtained under the zero-shot setting. To further explore the reasoning capability of SQL models, we investigate a few-shot prompting strategy, where several manually crafted Text-to-SQL exemplars, consisting of a natural language question and its corresponding SQL query, are incorporated into the system prompt to guide the model’s understanding of the mapping between language and database operations.

\begin{table}[htbp]
  \centering
  \caption{Few shot experiments.}
  \resizebox{0.8\linewidth}{!}{
    \begin{tabular}{l|ccc}
    \toprule
    \multicolumn{1}{c|}{\textbf{Model}} & \textbf{CA} & \textbf{EN} & \textbf{PA} \\
    \midrule
    \textbf{GPT-4o  + zero-shot} & 51.85 & 70.23 & 65.71 \\
    \textbf{ \qquad \quad + 1-shot} & 52.24 & 71.19 & 66.51 \\
    \textbf{ \qquad \quad+ 2-shot} & \underline{52.66} & \underline{71.58} & \underline{67.23} \\
    \textbf{ \qquad \quad+ 3-shot} & \textbf{53.84} & \textbf{72.75} & \textbf{67.46} \\
    \midrule
    \textbf{Qwen2.5-72B  + zero-shot} & 55.56 & 71.76 & 62.86 \\
    \textbf{ \qquad \quad+ 1-shot} & 56.43 & 72.62 & 63.70 \\
    \textbf{ \qquad \quad+ 2-shot} & \underline{57.16} & \underline{73.05} & \underline{64.21} \\
    \textbf{ \qquad \quad+ 3-shot} & \textbf{57.52} & \textbf{73.35} & \textbf{65.68} \\
    \bottomrule
    \end{tabular}%
    }
  \label{tab:few_shot}%
\end{table}%

As shown in Table~\ref{tab:few_shot}, both GPT-4o and Qwen2.5-72B-Instruct exhibit a monotonic improvement as the number of few-shot exemplars increases. For GPT-4o, the average score rises from 62.60 (zero-shot) to 64.68 (3-shot), yielding a relative gain of 3.3\%. A similar trend is observed for Qwen2.5-72B, which improves from 63.39 to 65.52 with 3-shot prompting, reflecting a relative gain of 3.4\%. These results confirm that few-shot prompting effectively helps the models adapt to the dynamic evaluation setting by leveraging in-context examples for contextual alignment.

However, the improvement plateaus beyond 2-shot, suggesting that the added examples primarily enhance pattern recall rather than strengthening the model’s underlying compositional reasoning. 
In conclusion, while few-shot prompting mitigates cold-start difficulties in dynamic Text-to-SQL interaction, it remains insufficient for handling deeper schema reasoning or cross-domain transfer. Future work may integrate schema-aware pretraining or reinforcement learning-based adaptation to further enhance model robustness under dynamic contexts.


\section{Conclusion}
In this work, we presented DySQL-Bench, the first benchmark designed to evaluate large language models under dynamic, multi-turn Text-to-SQL interaction scenarios. Unlike previous static datasets, DySQL-Bench captures the evolving nature of real-world database interactions where user intents, constraints, and analytical objectives continuously change over time. To construct this benchmark at scale and with high reliability, we proposed a two-stage automatic task synthesis and verification pipeline, which transforms raw database tables into structured tree representations and generates realistic interaction sequences through large language models, followed by interaction-oriented quality control and expert validation.

To enable realistic assessment, we further introduced a multi-turn user--model--database interaction framework that simulates natural dialogue between a user and a model in an executable database environment. This setting allows systematic evaluation of models’ ability to perform query reformulation, contextual reasoning, and adaptive error recovery across turns. Comprehensive experiments across 13 domains and 1,072 tasks, reveal that even frontier models such as GPT-4o, 
highlighting the substantial challenges of interactive SQL reasoning.
We hope DySQL-Bench will serve as a standardized and challenging platform for advancing research in dynamic Text-to-SQL, interactive data analysis, and context-aware reasoning, paving the way toward truly adaptive and conversational database intelligence.

\clearpage
\bibliographystyle{IEEEtran}
\bibliography{IEEEabrv,main.bib}

\end{document}